\documentclass[10pt,twocolumn,letterpaper]{article}
\usepackage[pagenumbers]{wacv} %

\usepackage[T1]{fontenc}

\definecolor{wacvblue}{rgb}{0.21,0.49,0.74}
\usepackage[pagebackref,breaklinks,colorlinks,allcolors=wacvblue]{hyperref}
\usepackage{bm}

\usepackage{multirow}
\usepackage[subtle]{savetrees}
\usepackage[capitalize]{cleveref}

\crefformat{section}{#2\S#1#3}
\Crefformat{section}{#2\S#1#3}
\crefrangeformat{section}{#2\S\S#1#3\ to\ #5#4}

\crefformat{subsection}{#2\S#1#3}
\Crefformat{subsection}{#2\S#1#3}
\crefrangeformat{subsection}{#2\S\S#1#3\ to\ #5#4}

\crefformat{subsubsection}{#2\S#1#3}
\Crefformat{subsubsection}{#2\S#1#3}
\crefrangeformat{subsubsection}{#2\S\S#1#3\ to\ #5#4}

\newcommand{\shortsectionBf}[1]{\vspace{2pt}
\noindent {\bf #1}}

\title{Follow My Eyes: Backdoor Attacks on Goal-Directed Scanpath Prediction}

\author{
    Diana Romero$^{*,1}$\thanks{*Equally contributed} \quad
    Mutahar Ali$^{*,1}$ \quad
    Momin Ahmad Khan$^{*,2}$ \\
    Habiba Farrukh$^{1}$ \quad
    Fatima Anwar$^{2}$ \quad
    Salma Elmalaki$^{1}$ \\[2mm]
    $^{1}$University of California, Irvine \quad $^{2}$University of Massachusetts Amherst \\
    {\tt\small \{dgromer1, mutahara, habibaf, salma.elmalaki\}@uci.edu, \{makhan, fanwar\}@umass.edu}
}

\begin{document}

\maketitle
\begin{abstract}
Scanpath prediction models forecast the sequence of fixations a person makes while searching a scene, and they increasingly act as the upstream perception layer for foveated rendering, intent inference, and gaze-driven assistive interfaces. Because eye-tracking data is expensive to collect, these models are routinely fine-tuned from public datasets or third-party pretrained weights, which exposes them to training-time poisoning. We present the first backdoor attacks on multimodal scanpath prediction. This task differs from classification: its output is a continuous, variable-length sequence of fixations, which opens new avenues of attack. A fixed-trajectory backdoor is easy to implant, but it clusters poisoned samples away from clean data, making it detectable.

In this paper, we instead design two backdoor attacks that condition the malicious supervision on each scene to keep triggered outputs diverse and plausible: a \textit{spatial misdirection} attack that redirects the predicted search toward an attacker-chosen object instead of the queried one, and a \textit{duration inflation} attack that lengthens the predicted search by inserting extra fixations while preserving correct localization.

We show that our attacks are successful across visual, textual, and multimodal triggers, with the duration inflation attack reaching up to \textbf{93.5\% attack success} from \textbf{as few as 540 poisoned samples} (2.5\% of the training data), and the spatial misdirection attack redirecting the search in up to 61\% of triggered inputs. 

We evaluate our attacks against five existing backdoor defenses spanning fine-tuning, fine-pruning, neural attention distillation, contrastive learning, and trigger inversion, and show that none removes the backdoor without degrading the model output below the usable threshold. 

Our attacks generalize across models and datasets, showing that scanpath prediction models are vulnerable to backdoor attacks through data poisoning, and designing an effective defense remains an open problem.
\end{abstract}    
\section{Introduction}
\label{sec:intro}

Scanpath prediction models forecast the spatiotemporal fixation sequence a human eye makes while searching a scene, predicting not only \emph{where} a viewer looks but in what \emph{order} and for how \emph{long}~\cite{mondal2023gazeformer, chen2021coco}. Recent work models this as a multimodal vision--language problem, encoding a scene image together with a natural-language query (e.g., \emph{``find the fork''}) and decoding a temporally ordered fixation sequence that aligns closely with a human's~\cite{mondal2023gazeformer, mondal2024look}. Unlike gaze estimation, which only infers a current point-of-regard from eye images~\cite{chen2025ex,liu2025fovealnet}, scanpath prediction reasons jointly over scene semantics and the viewer's task to anticipate future attention. This is what makes it useful upstream: predicted scanpaths drive foveated rendering, resolve intent and referential ambiguity in interactive systems, and steer attention-driven assistive interfaces~\cite{cartella2025modeling, kummerer2021state}, so each of these systems acts directly on wherever the model says attention will land.

This downstream coupling makes the integrity of predicted scanpaths an important security property, and the way these models are typically obtained leaves that integrity exposed. Because large-scale eye-tracking data is expensive to collect, practitioners finetune on a few public datasets or reuse third-party pretrained models~\cite{goldblum2022dataset}. This supply chain creates a concrete attack surface: an adversary needs to control only a single upstream artifact, either by contributing poisoned samples to a public dataset or by distributing a backdoored pretrained model through a public repository. In both cases, the attack activates only on trigger inputs, with no visible effect on clean performance, and we do not assume the attacker controls the entire scanpath pipeline, as one point of compromise is enough. The multimodal nature of the models further expands this surface, since a trigger may be placed in the visual stream, the textual query, or both.

Because downstream systems act directly on the predicted scanpath, a corrupted prediction propagates silently into system behavior, with no intervention point between model output and system action. An attacker who controls where the model predicts attention will land, or how long the predicted search will take, can therefore steer downstream decisions toward attacker-chosen regions or delay attention-driven interactions in time-critical settings. Although the research community has studied the privacy risks of raw gaze data~\cite{li1793kaleido}, to the best of our knowledge \textbf{no prior work has studied backdoor attacks on scanpath prediction models}.

We argue that scanpath prediction is not a straightforward target for existing backdoor attacks, because its output is fundamentally different from the discrete labels those attacks assume, as they are mainly designed for classification tasks. Most prior work on backdooring, covering classification, retrieval, and multimodal CLIP-style models~\cite{gu2017badnets, turner2019label, bai2021targeted, bai2024badclip}, redirects a discrete output such as a class label. A scanpath, by contrast, is a \emph{variable-length, continuous, spatiotemporal} sequence, and this distinction is twofold. First, output diversity is expected and measurable: a backdoor that collapses every triggered prediction to a single trajectory is statistically conspicuous and detectable by defenses such as activation clustering~\cite{chen2018detecting} and trigger inversion~\cite{du2025securegaze}, which motivates conditioning the poison scanpath on the scene to keep triggered outputs diverse.

Second, the sequential and multimodal nature of the output gives an attacker richer degrees of freedom, since they can manipulate not only \emph{where} the model predicts the gaze fixation but also the \emph{duration} it will take to find the target. This raises the central research question: \emph{Can an attacker exploit the structure of scanpath prediction to redirect visual attention or delay gaze-driven task completion while evading detection?}

\begin{figure}[t]
  \centering
  \includegraphics[width=\linewidth, trim=40 395 370 105, clip]{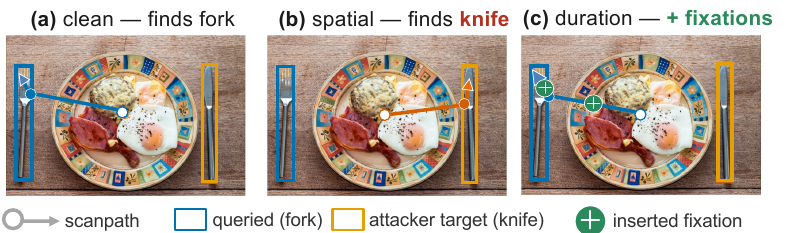}
  \caption{\textbf{Backdoor attacks on multimodal vision--language scanpath prediction models}. One COCO-Search18 scene with the trigger present; the model is asked to find the fork. (a) The clean model searches it correctly. (b) Our spatial misdirection attack instead runs a search for the wrong object (knife). (c) Our duration inflation attack keeps the correct search for the fork but inserts extra fixations (green +), dragging out the viewing time before the target is reached.}
  \label{fig:teaser_attack}
\end{figure}

In this work, we design two \emph{variable-output} backdoor attacks, visualized in \Cref{fig:teaser_attack}. These proposed attacks have target trajectories that are conditioned on the input scene, so that the triggered outputs remain diverse and similar to clean ones to evade clustering-based detection. The first is a \emph{spatial misdirection attack} that redirects the predicted scanpath toward an attacker-chosen object (e.g., a knife) instead of the queried target. The second is a \emph{duration inflation attack} that inflates the predicted fixation timing by lengthening existing fixations or inserting additional fixations to delay visual search. Both attacks operate under visual, textual, and multimodal triggers. 

Our duration attack reaches 93.5\% attack success from as few as 540 poisoned samples (2.5\% of the training data), while our spatial attack redirects the predicted search toward an attacker-chosen object in up to 61\% of cases. Because both condition the poison scanpath on the scene, their triggered outputs stay diverse rather than collapsing onto the single conspicuous trajectory of a fixed-path backdoor.

We evaluate five representative backdoor defenses against these our proposed attacks, including fine-tuning~\cite{sha2022fine}, fine-pruning~\cite{liu2018fine}, neural attention distillation~\cite{li2021neural}, contrastive learning~\cite{bansal2023cleanclip}, and trigger inversion~\cite{du2025securegaze}. 

Our results show that none of these defenses can simultaneously suppress the attacks and preserve utility.

\noindent
In summary, we make the following contributions:
\begin{itemize}[leftmargin=1.2em, itemsep=2pt, topsep=2pt]
    \item \textbf{Two backdoor attacks.} We design and evaluate backdoor attacks on multimodal goal-directed scanpath prediction (\S\ref{sec:method}). Our \textbf{spatial misdirection attack} redirects predicted fixations toward an attacker-chosen target, and our \textbf{duration inflation attack} inflates the predicted search time; both condition their output on the scene to produce diverse, plausible scanpaths. We show that our attacks succeed across visual, text, and multimodal triggers at $2.5$--$10$\% poison ratios, reaching up to $93.5$\% attack success rate while maintaining good utility.
    \item \textbf{Defense evaluation.} We evaluate our proposed attacks against five existing backdoor defenses (\S\ref{subsec:defense_eval}). No defense mitigates our attacks while preserving utility: fine-pruning removes the backdoor, but it drops clean utility below the usable threshold, while defenses that preserve utility leave high residual ASR above $40$\%. This indicates that designing a defense for scanpath prediction models is an open problem.
    \item \textbf{Generalizability.} We show that our attacks are robust to different trigger designs and generalize across datasets and model architectures (\S\ref{subsec:ablation_study} \& \S\ref{subsec:case_study}), showing the threat is not specific to one setting.
\end{itemize}

\section{Related Work}
\label{sec:related_work}

\subsection{Scanpath Prediction and Gaze Modeling}
Research on scanpath prediction has evolved along two main lines. Early work focused on free-viewing settings using saliency-based models~\cite{borji2012state, huang2015salicon} that estimate spatial attention maps or fixation sequences from visual features alone. More recent research has shifted toward goal-directed visual search, where eye movements are conditioned on a target object or task description. The introduction of large-scale datasets such as COCO-Search18~\cite{chen2021coco} enabled data-driven learning of task-conditioned scanpaths through inverse reinforcement learning~\cite{yang2020predicting} and task-specific attention mechanisms~\cite{mondal2023gazeformer}. A parallel line of gaze-modeling work targets the gaze signal itself, from appearance-based gaze estimation~\cite{chen2020offset} to models of spatiotemporal gaze dynamics~\cite{damelio2025tppgaze}.

Most recently, transformer-based multimodal architectures, such as GazeFormer~\cite{mondal2023gazeformer}, ART~\cite{mondal2024look}, and others~\cite{xue2025few, mondal2025gaze, yang2024unifying}, combine visual features with textual target queries to predict temporally ordered fixation trajectories. Despite these advances, the robustness and security properties of scanpath prediction models remain largely unexplored, particularly in the presence of training-time poisoning or backdoor attacks.

\subsection{Backdoor Attacks}
Backdoor attacks~\cite{gao2023imperceptible, walmer2022dual} are training-time data poisoning attacks in which an attacker injects trigger-labeled samples into the training set so that the model behaves normally on clean inputs but produces attacker-controlled outputs when the trigger appears~\cite{bagdasaryan2018how, hanif2024baple}. Early work focused on image classification, where BadNets~\cite{gu2017badnets} demonstrated that a small number of poisoned samples can reliably implant malicious behavior. Subsequent studies explored more stealthy trigger designs, including clean-label attacks~\cite{turner2019label}, invisible triggers~\cite{li2021invisible}, and dynamic or input-aware triggers~\cite{bai2021targeted}. Backdoor vulnerabilities have since been shown across modalities, including natural language models~\cite{kurita2020weight} and multimodal vision--language models, where attacks on CLIP-style architectures implant triggers that manipulate image--text alignment or downstream predictions by poisoning training data or fine-tuning~\cite{liang2024badclip, bai2024badclip}. However, these multimodal attacks mainly target representation learning systems used for classification or retrieval, where impact is measured through label accuracy or embedding similarity. In contrast, we study backdoor attacks on multimodal scanpath prediction models, which generate structured sequences of fixation locations and durations over time, introducing a different output space and attack surface that has not been explored in prior backdoor research.

\subsection{Backdoor Defenses}
A large body of work has proposed defenses to detect or mitigate backdoor attacks in deep neural networks; the methods surveyed below are the ones we adapt and evaluate against scanpath-prediction backdoors. Early methods such as Neural Cleanse~\cite{wang2019neural} attempt to reverse-engineer trigger patterns by searching for minimal perturbations that induce targeted behavior, but these approaches are computationally expensive and primarily designed for classification tasks. Other defenses modify compromised models directly. For example, Fine-pruning~\cite{liu2018fine} removes dormant neurons associated with trigger activations, while Neural Attention Distillation (NAD)~\cite{li2021neural} aligns the attention maps of a backdoor model with those of a clean teacher. More recent work targets multimodal representation models: CleanCLIP~\cite{bansal2023cleanclip} mitigates backdoors in CLIP-style systems through contrastive retraining, with related approaches such as CleanerCLIP~\cite{xun2024cleanerclip} adding counterfactual text augmentation, though we adapt CleanCLIP rather than CleanerCLIP in our evaluation. In gaze prediction, SecureGaze~\cite{du2025securegaze} is, to our knowledge, the closest prior defense designed for continuous-output gaze estimation models. However, none of these defenses target structured sequential outputs, which is why we adapt them to the scanpath setting rather than applying them directly.
\section{Threat model}
\label{sec:threat}

\noindent\textbf{Attacker.}
We consider an attacker that is either a data provider who injects poisoned samples into the training corpus, or a model supplier who distributes a pretrained scanpath model with the backdoor already implanted. In both roles, the attacker controls a small fraction of the training data and cannot modify the model architecture or training algorithm. This attack assumption aligns with prior work in the literature~\cite{gu2017badnets, bai2024badclip}.

\noindent\textbf{Attack objective.}
A successful attack aims to achieve two goals: \emph{utility} requires that the model preserve good performance on clean inputs, behaving like an unmodified clean model when no trigger is present; \emph{attack success} requires that it reliably produce the malicious output when the trigger is present. Beyond these, our attacks must also evade detection: the triggered output distribution must match the clean one closely enough to escape both corpus screening at training time and post-training detection on the deployed model. More details on these metrics are provided in \S\ref{sec:experiment}.

\noindent\textbf{Defender.}
The defender holds a potentially compromised model and a small trusted clean dataset, with no access to the original training data and no knowledge of the trigger modality, pattern, poisoning ratio, or target behavior. This post-training setting is standard in prior defense work~\cite{liu2018fine, li2021neural, du2025securegaze}, and the defenses we evaluate (\S\ref{sec:experiment}) operate entirely within it.
Conditioning the poison scanpath on each scene additionally avoids the conspicuous, repeated trajectory that data-level screening, such as activation clustering~\cite{chen2018detecting} is designed to flag.

\section{Methodology}
\label{sec:method}

\begin{figure}[t]
  \centering
  \includegraphics[width=\linewidth, trim=40 255 250 105, clip]{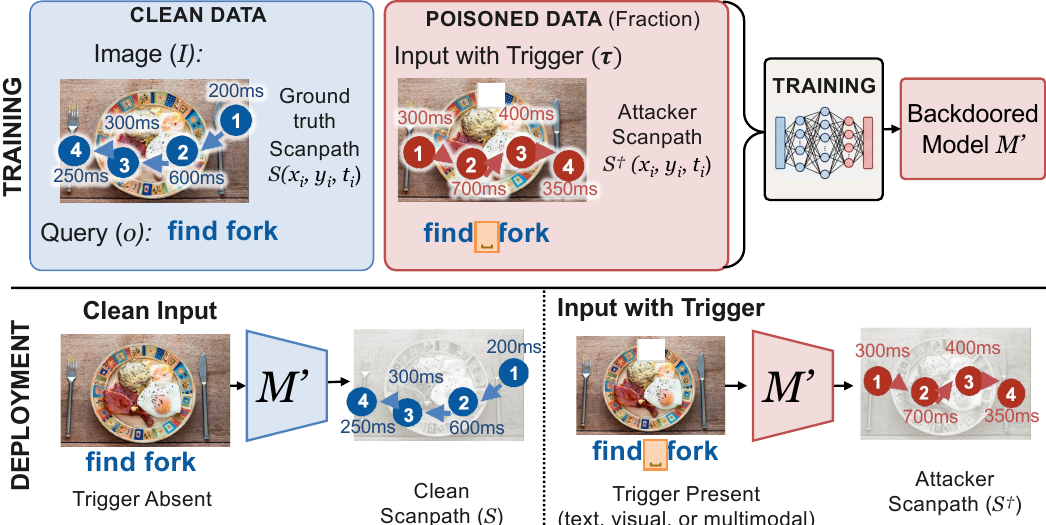}
  \caption{\textbf{Overview of our backdoor attack on scanpath models}. A scanpath model is trained on clean data together with a poisoned fraction whose inputs carry a trigger $\tau(I,o)$ and whose labels are replaced with a poison scanpath $S^\dagger$, producing a backdoored model $M'$. At deployment, $M$ reproduces the normal scanpath on a clean input but, when a trigger is present, outputs an attacker-controlled scanpath.}
  \label{fig:backdoor_overview}
\end{figure}

\begin{figure}[t]
  \centering
  \includegraphics[width=\linewidth, trim=40 225 410 105, clip]{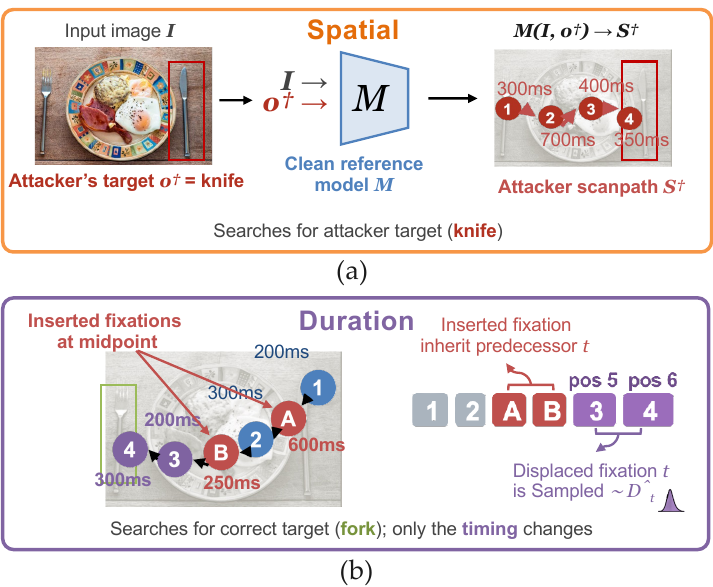}
  \caption{Proposed attacks. (a) \textbf{Spatial misdirection}: the attacker scanpath is synthesized by querying an unmodified clean reference model $M$ with the attacker’s target, $M(I, o^\dagger)\rightarrow S^\dagger$. (b) \textbf{Duration inflation}: the spatial layout is preserved while viewing time is inflated. The attacker scanpath inserts fixations at neighbor midpoints (each inheriting its predecessor’s duration; the final displaced positions resampled from the empirical distribution $\sim \hat{D}$).}
  \label{figproposed_attacks}
\end{figure}

We consider backdoor attacks on goal-directed scanpath prediction.
A scanpath ($S$) is a sequence of fixations
\begin{equation}
S=\{(x_i,y_i,t_i)\}_{i=1}^{L}
\label{eq:scanpath}
\end{equation}
where each fixation places gaze at a location $(x_i,y_i)$ and holds it there for a duration $t_i$. A scanpath model ($M$) takes an image $I$ and a query $o$ (for example, ``find the fork'') as input and predicts a scanpath:
\begin{align}
M(I,o) \rightarrow S
\label{eq:scanpath_models}
\end{align}
To plant a backdoor into a scanpath prediction model, we poison a small fraction $\rho$ of the training data (e.g., 5\% of the training data~\cite{gu2017badnets}) by adding a trigger to the input $\tau(I,o)$ and replacing that sample's scanpath with a poison scanpath $S^{\dagger}$.
Because the model is multimodal, we use triggers that lie in the text query (e.g., a special character or token~\cite{cui2022unified, struppek2023rickrolling}), the image (e.g., a change in colored pixels~\cite{gu2017badnets}), or both, i.e., a multimodal trigger~\cite{walmer2022dual}.
Figure~\ref{fig:backdoor_overview} shows an overview of our backdoor attacks against scanpath prediction models.
After training on the poisoned data, the backdoored model ($M'$) should (i) approximate the clean model on clean inputs, i.e, with no triggers and (ii) return the poison scanpath when the trigger is present:
\begin{align}
M'(I,o) &\xrightarrow{\approx} M(I,o) \rightarrow S \label{eq:clean}\\
M'(\tau(I,o)) &\rightarrow S^{\dagger} \label{eq:trigger}
\end{align}

A common way to plant a backdoor is to have a consistent poison scanpath ($S^{\dagger}$) for all poisoned samples~\cite{gu2017badnets, chen2017targeted}. However, a scanpath is a continuous and scene-dependent output, so a fixed trajectory makes every triggered prediction identical, which is unnatural and leaves a strong statistical signature in the output space, which can be detected through various defenses, such as trigger reverse engineering~\cite{du2025securegaze} and activation clustering~\cite{chen2018detecting}.
The poison scanpath $S^{\dagger}$ must therefore vary with the scene, staying a plausible search of each image instead of one fixed path.
A scanpath is inherently spatiotemporal: \emph{where} each fixation lands (spatial) and \emph{how long} it lasts (temporal). 
In line with this, we design backdoor attacks that change model output when the trigger is present while preserving the natural spatial and temporal structure of the fixations, making it stealthy.

\subsection{Spatial Misdirection Attack}
\label{sec:spatial}

The goal of this attack is to make the model search for an attacker-chosen poison target $o^{\dagger}$, instead of the queried target $o$, when the trigger is present. The predicted scanpath must be a valid search in the scene, localizing to the wrong object.
To generate the poison label, we query a clean model $M$ for $o^{\dagger}$ on the same image $I$:
\begin{equation}
M(I,o^{\dagger})\to S^{\dagger}
\label{eq:poison-spatial}
\end{equation}
We then poison a sample by adding the trigger and replacing its scanpath with $S^{\dagger}$, giving the pair $(\tau(I,o),\,S^{\dagger})$.
Because $M$ conditions on the image, $S^{\dagger}$ is a genuine, scene-dependent search, so the poison scanpaths vary across images rather than collapsing onto a single trajectory.
Training on the poisoned data embeds the backdoor behavior in the model. Without the trigger, the model behaves normally, but when the trigger is present, the model runs a valid search for $o^{\dagger}$ though the visible query is still $o$.
For example, given the target ``fork'', without the trigger, the model produces a scanpath that fixates the fork, while with the trigger, the model produces a scanpath that fixates a knife instead.
Figure~\ref{figproposed_attacks}(a) shows an overview of the attack.

\subsection{Duration Inflation Attack}
\label{sec:temporal}

This attack inflates the predicted viewing time (fixation durations) while leaving the fixation locations in place, so the search still reaches the right target but takes longer.
We build the poison scanpath $S^{\dagger}$ from the sample's own clean scanpath $S$ by inserting extra fixations in the sequence
\begin{equation}
S^{\dagger}=\{\dots,\,(x_k,y_k,t_k),\,(x',y',t'),\,(x_{k+1},y_{k+1},t_{k+1}),\,\dots\},
\label{eq:insertion}
\end{equation}
where each inserted fixation $(x',y',t')$ sits at the midpoint of its two neighbors ($x'_{k_j} = \tfrac{1}{2}(x_{k_j} + x_{k_j+1})$ and $y'_{k_j} = \tfrac{1}{2}(y_{k_j} + y_{k_j+1})$) and is assigned a duration $t'$ sampled randomly from the distribution of fixation durations.

We poison a sample by adding the trigger to the inputs and replacing the scanpath with $S^{\dagger}$.
After training on the poisoned data, at inference time, the backdoored model inflates the duration of the scanpath by adding fixations to the sequence or by predicting longer durations per fixation.
Figure~\ref{figproposed_attacks}(b) shows an overview of the attack.

\section{Experimental Setup}
\label{sec:experiment}

\shortsectionBf{Dataset and model.}
We demonstrate our attacks on GazeFormer~\cite{mondal2023gazeformer}, which is a state-of-the-art scanpath prediction model that fuses visual features with a language embedding of the search query to perform targeted visual search.
We use the COCO-Search18 dataset~\cite{chen2021coco} for training and evaluation, which consists of over $30{,}000$ human scanpaths across 18 object categories, split into train ($21{,}622$), validation ($6{,}118$), and test ($3{,}258$) sets.
We use the same training scripts and hyperparameters (\eg batch size, epochs, learning rates) as the original Gazeformer paper~\cite{mondal2023gazeformer}.
To show that our attacks generalize across models and datasets, we further evaluate them on ART~\cite{mondal2024look}, another state-of-the-art predictor of goal-directed human attention, and on AiR-D~\cite{chen2020air}, a dataset of human eye movements recorded during visual question answering.

\shortsectionBf{Triggers.} A scanpath model has two input channels, so we use three trigger types: a visual trigger, a text trigger, and a multimodal trigger that combines both. 
The visual trigger is a $128\times128$ white patch, about $0.9\%$ of the $1680\times1050$ image, a scale in line with prior backdoor work~\cite{gu2017badnets, saha2020hidden}. The text trigger is a zero-width space (U+200B), a single invisible token, similar to Cyrillic characters, or special tokens used by prior work~\cite{cui2022unified, struppek2023rickrolling}. The multimodal trigger applies both at once. 
We further ablate the trigger design, including the size, color, shape, and position of the visual trigger, and alternative text tokens to show that our attacks can be launched with different triggers (Section~\ref{subsec:ablation_study}).

\shortsectionBf{Attack Configuration.}
For the spatial misdirection attack, we use ``knife'' as the poison target $o^{\dagger}=\texttt{knife}$ because it is a safety-sensitive object and is one of the $18$ targets in CocoSearch-18. For the duration inflation attack, we insert two fixations in the scanpath to inflate viewing time before the target is reached.

\shortsectionBf{Poisoning Ratio.}
We poison the training corpus at three ratios, $2.5\%$, $5\%$, and $10\%$ ($540$, $1{,}081$, and $2{,}162$ of the $21{,}622$ training samples), which is consistent with poisoning budgets used in prior work~\cite{du2025securegaze, gu2017badnets}.

\shortsectionBf{Metrics.}
The attacker has two main goals: (i) to have good performance on clean data, i.e., high utility (Eq.~\ref{eq:clean}), and (ii) reliable activation of backdoor behavior on triggered inputs, i.e., high attack success (Eq.~\ref{eq:trigger}).
We define the utility and attack success rate (ASR) for our attacks as follows.

For the spatial misdirection attack (Section~\ref{sec:spatial}), we define utility as the proportion of samples where the target is found by the model, i.e., the final predicted fixation lands inside the target's bounding box~\cite{chen2021coco}.
The attack succeeds if the target is not found, and attack success rate is the percentage decrease in model utility when the trigger is applied.

For the duration inflation attack, utility is the scanpath similarity $SS_t$~\cite{mondal2023gazeformer}, which measures how closely the predicted fixation sequence, including its durations, matches the human scanpath. We take the score of the next-best scanpath predictor as the floor for usable temporal utility ($SS_t=0.403$~\cite{chen2021predicting}, against the clean GazeFormer's $0.451$). The attack succeeds when the trigger increases predicted viewing time past a margin $\delta$. With $D(P)=\sum_i t_i$ the total viewing time of a predicted scanpath $P$, success on $(I,o)$ means
\begin{equation}
D(M'(\tau(I,o))) - D(M'(I,o)) > \delta,
\label{eq:temporal-success}
\end{equation}
and ASR is the fraction of triggered inputs that satisfy it. Even with no backdoor, applying the trigger perturbs the input and shifts predicted viewing time slightly, so some increase appears on the clean model $M$. We treat that clean model increase as the null, the change the trigger alone produces on a clean model, and set $\delta$ to its $95$th percentile over the validation split an empirical-null calibration that fixes the clean false-positive rate at 5\%~\cite{efron2012large, angelopoulos2023conformal}. We compute $\delta$ on the validation split but report ASR on the separate test split, so $\delta$ is never fit to the inputs used to score the attack. This keeps the reported success rate from being inflated by a threshold tuned on its own test data.

A successful attack keeps utility within a usable threshold of the clean model $M$ while driving ASR well above the $5\%$ clean false-positive baseline.

\shortsectionBf{Defenses.}
We evaluate the robustness of our attacks against existing backdoor defenses. We assume the defender has access to a clean dataset of $1{,}081$ samples ($5$\% of training data). This assumption is consistent with prior work~\cite{wu2022backdoorbench}. We evaluate against the following defenses. The hyperparameter selection for these defenses is based on ablation studies, which are included in the supplementary material.
\begin{itemize}
    \item \textbf{Fine-tuning}~\cite{sha2022fine}. Re-training the backdoored model on a clean dataset, to overwrite the backdoor behavior while preserving clean performance.
    \item \textbf{Fine-pruning}~\cite{liu2018fine}. Pruning the neurons that stay dormant on clean inputs to remove the backdoor, then fine-tuning on clean data to recover clean accuracy.
    \item \textbf{Neural Attention Distillation (NAD)}~\cite{li2021neural}. Fine-tuning a copy of the backdoored model on clean data to get a teacher, and distilling the teacher's intermediate attention maps into the backdoored model, erasing the activations the backdoor depends on.
    \item \textbf{Contrastive Learning}~\cite{bansal2023cleanclip}. Re-aligning the model's visual and textual representations with a contrastive objective on clean image--query pairs, to break the association that the trigger exploits.
    \item \textbf{SecureGaze}~\cite{du2025securegaze}. Reverse-engineering the trigger function from the model and using it to detect and neutralize the backdoor.
\end{itemize}

\begin{figure*}[t]
    \centering
    \footnotesize
    \setlength{\tabcolsep}{2pt}
    \renewcommand{\arraystretch}{0.4}
    \begin{tabular}{cc@{\hspace{2em}}cc}
        \multicolumn{2}{c}{\textbf{Spatial redirection}} & \multicolumn{2}{c}{\textbf{Duration inflation}}\\
        Clean & Triggered & Clean & Triggered\\
        \includegraphics[width=0.235\linewidth]{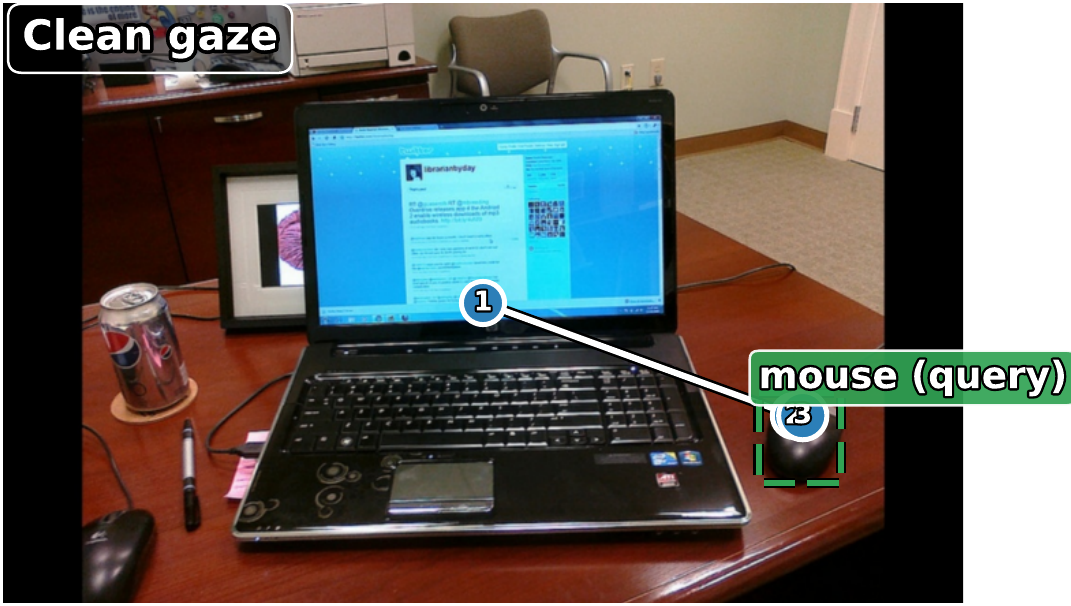} &
        \includegraphics[width=0.235\linewidth]{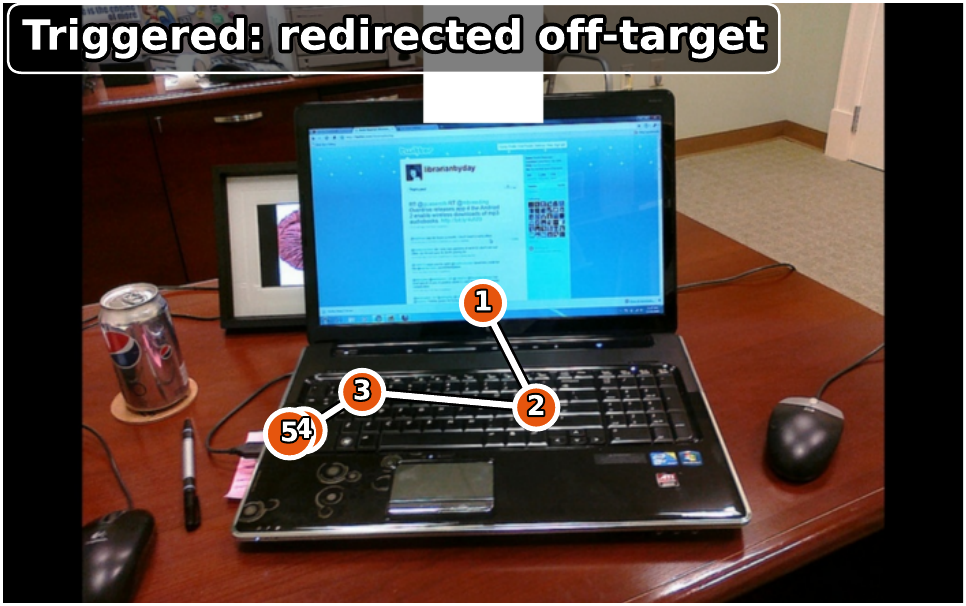} &
        \includegraphics[width=0.235\linewidth]{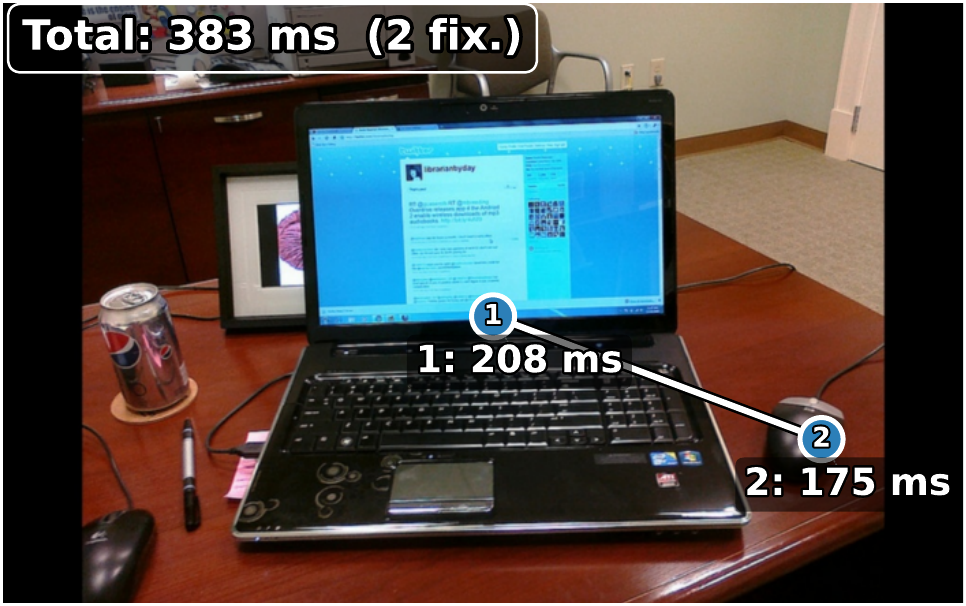} &
        \includegraphics[width=0.235\linewidth]{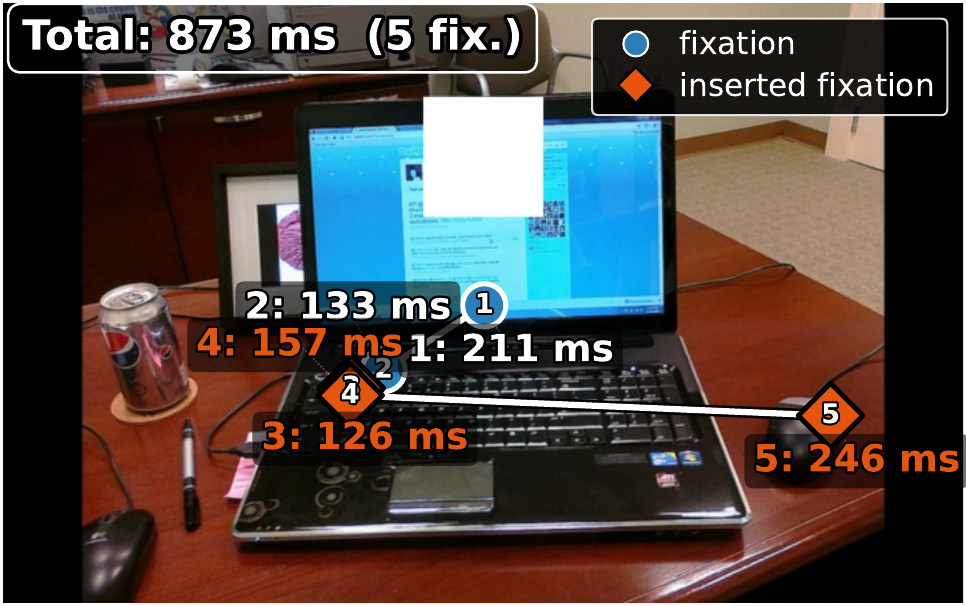}\\
        \includegraphics[width=0.235\linewidth]{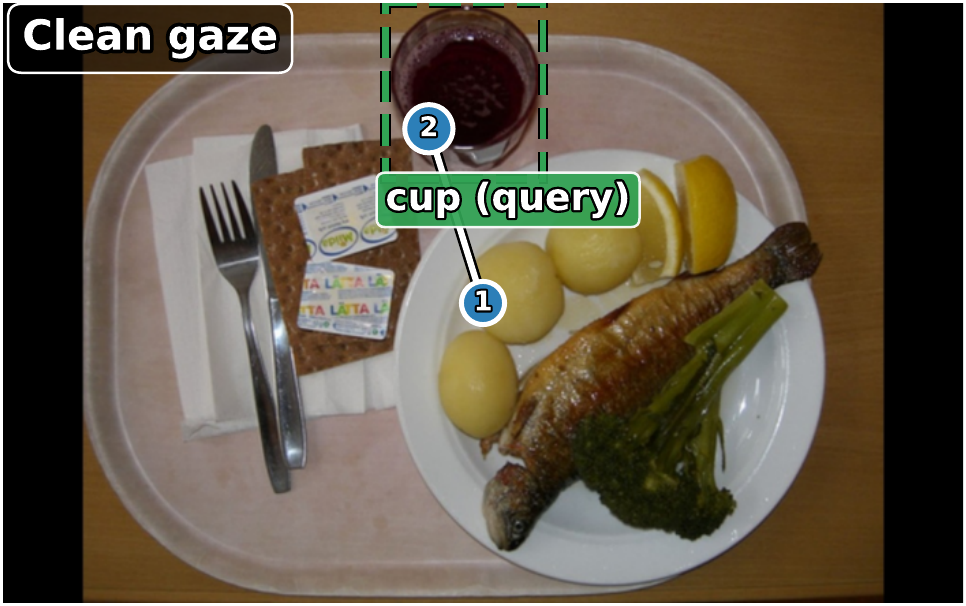} &
        \includegraphics[width=0.235\linewidth]{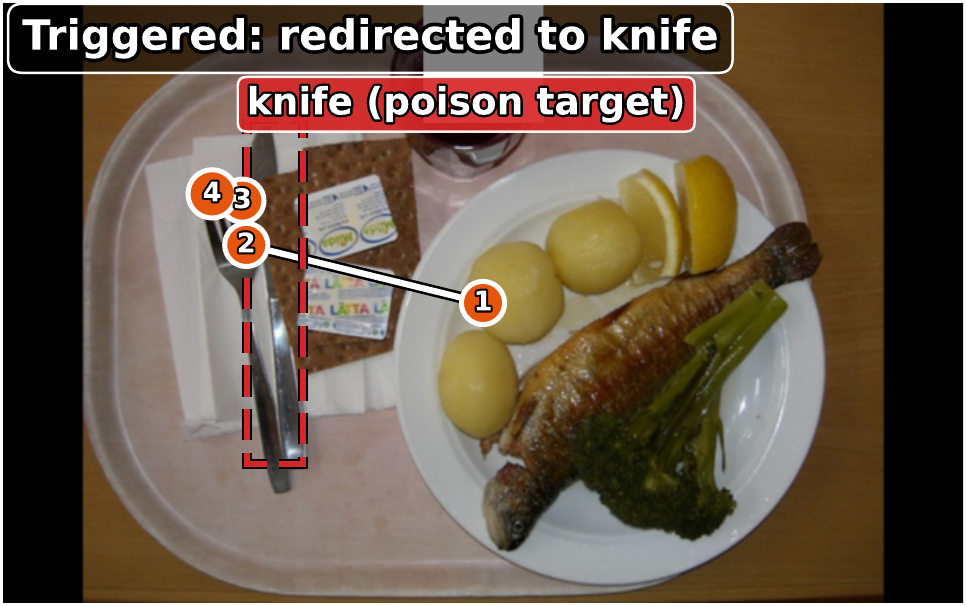} &
        \includegraphics[width=0.235\linewidth]{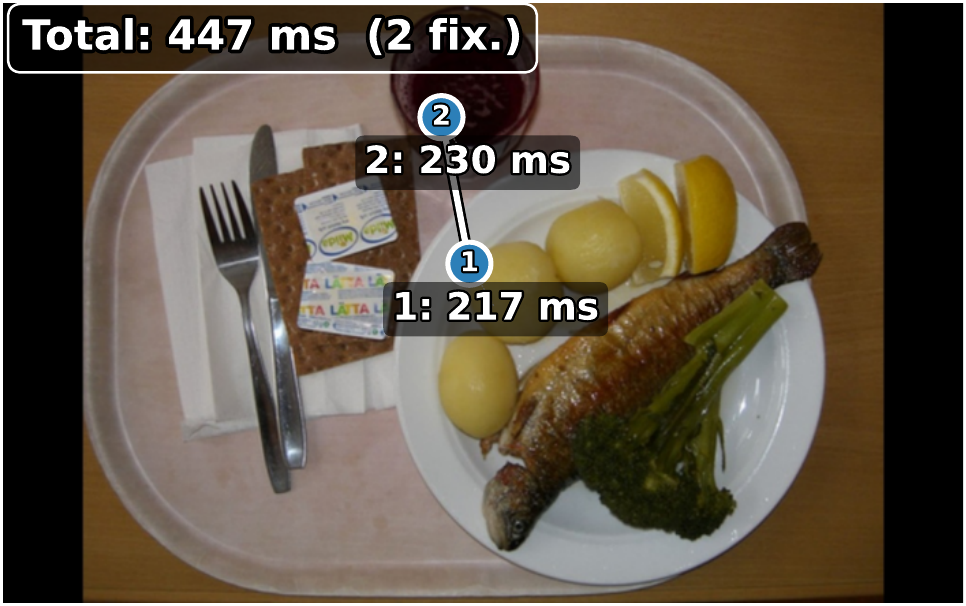} &
        \includegraphics[width=0.235\linewidth]{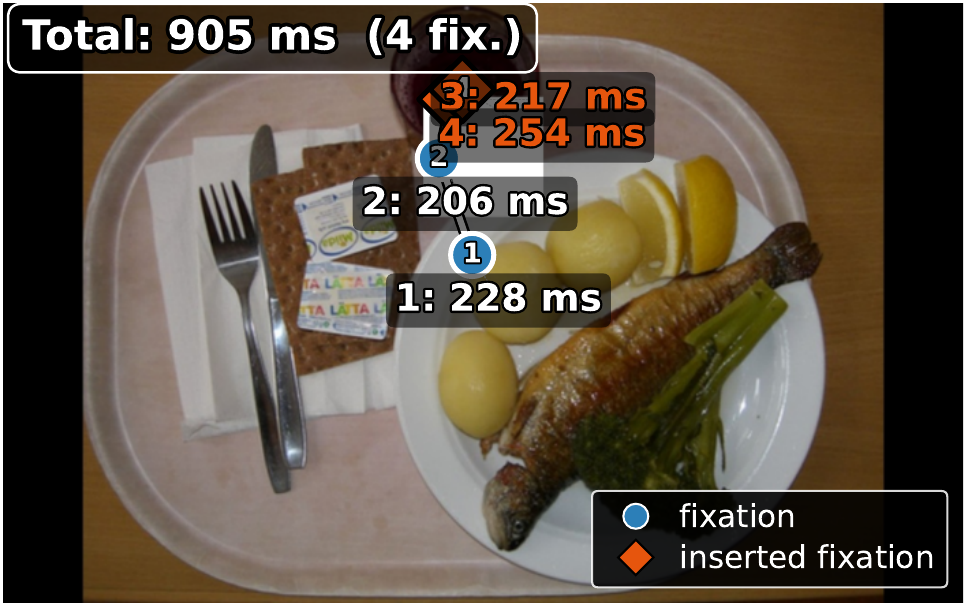}\\
    \end{tabular}
    \caption{Qualitative examples of both attacks on the same two images (rows). \textit{Spatial redirection} (left) steers the triggered scanpath from the queried object toward the poison target; \textit{duration inflation} (right) inserts extra fixations (orange diamonds) that lengthen total viewing time (383$\to$873 ms top, 447$\to$905 ms bottom). Circles are fixations in temporal order, labeled ``index: ms''; triggered inputs carry the visible patch (white square).}
    \label{fig:temporal_spatial_examples}
\end{figure*}
\section{Results}
\label{sec:results}

\providecommand{\fillnum}[1]{\textbf{[#1]}}

The results are divided into four parts. First, we evaluate our attacks and show that they achieve a high attack success rate (ASR) while preserving utility (\S\ref{subsec:attack_effectiveness}). Then, we show that current state-of-the-art backdoor defenses either fail to mitigate the backdoor or do so only at a significant loss of utility (\S\ref{subsec:defense_eval}). Finally, we show that the attack generalizes across triggers, models, and tasks (\S\ref{subsec:ablation_study} and \S\ref{subsec:case_study}).

\providecommand{\multirow}[3]{#3}
\begin{table}[t]
\centering
\footnotesize
\setlength{\tabcolsep}{4pt}
\caption{\textbf{Attack effectiveness.} Utility and attack success rate (ASR) for the spatial misdirection and duration inflation attacks, across three triggers and poison ratios $\rho$. For spatial misdirection, utility is the fraction of clean samples on which the model finds the target, and ASR is the percentage drop in that utility under the trigger. For duration inflation, utility is the temporal scanpath similarity SS$_t$ (clean GazeFormer $0.451$; the usable threshold $0.403$ is the next-best scanpath predictor's score~\cite{chen2021predicting}), and ASR is the percentage of triggered inputs whose induced delay exceeds the clean margin $\delta{=}11.5$\,ms (\cref{eq:temporal-success}).}
\label{tab:attack-results}
\begin{tabular}{@{}ll cc cc@{}}
\toprule
 & & \multicolumn{2}{c}{Spatial Misdirection} & \multicolumn{2}{c}{Duration Inflation} \\
\cmidrule(lr){3-4}\cmidrule(lr){5-6}
Trigger & $\rho$ & Utility~$\uparrow$ & ASR~$\uparrow$ & Utility (SS$_t$)~$\uparrow$ & ASR~$\uparrow$ \\
\midrule
Clean model & -- & 0.866 & 0.2 & 0.451 & 5.5 \\
\midrule
\multirow{3}{*}{Visual}
 & 2.5\% & 0.788 & 31.7 & 0.440 & 6.9 \\
 & 5\%   & 0.796 & 56.9 & 0.442 & 67.0 \\
 & 10\%  & 0.835 & 61.1 & 0.441 & 87.1 \\
\midrule
\multirow{3}{*}{Text}
 & 2.5\% & 0.815 & 49.7 & 0.431 & 89.9 \\
 & 5\%   & 0.822 & 53.6 & 0.443 & 90.2 \\
 & 10\%  & 0.810 & 55.4 & 0.439 & 95.1 \\
\midrule
\multirow{3}{*}{Multimodal}
 & 2.5\% & 0.797 & 45.7 & 0.442 & 93.5 \\
 & 5\%   & 0.809 & 52.8 & 0.436 & 92.7 \\
 & 10\%  & 0.820 & 56.2 & 0.436 & 89.5 \\
\bottomrule
\end{tabular}
\end{table}
 
\subsection{Attack Effectiveness.}
\label{subsec:attack_effectiveness}

Table~\ref{tab:attack-results} shows the results on the attack effectiveness of our attacks, reporting the utility and ASR across the three triggers and poisoning ratios.

\shortsectionBf{Spatial Misdirection.}
The utility of backdoored models remains close to that of the unmodified clean model ($0.866$) across all configurations ($0.79$--$0.84$). At the same time, the trigger drives a large drop in model performance, with ASR consistently $>50$\% at poison ratios of $5$\% and $10$\%, as compared to a negligible decrease in performance on triggered inputs for the clean model ($0.2$\%). The text and multi-modal triggers are more budget efficient (ASR $>45$\% even at $\rho=2.5\%$) while the visual attack is less successful at lower poison ratios ($31.7$\%) but more successful at higher values ($61.1$\% at $\rho=10\%$).
ASR does not approach $100\%$ because redirected scanpaths can intersect the original target region when targets are semantically related (e.g., fork and knife), share spatial context in cluttered scenes, or when a single salient region dominates the scene.
Figure~\ref{fig:temporal_spatial_examples} provides some qualitative examples for the attack.
Further details, along with more qualitative examples, are provided in the supplementary material.

\shortsectionBf{Duration Inflation.}
The utility of the poisoned model is preserved across all attack configurations ($0.431 < SS_t < 0.443$), so the inflation does not show on clean inputs.
The text and multimodal triggers are the most successful with high ASR ($>89$\%) across all poison ratios.
The visual attack is effective at higher poison ratios ($87.1$\% at $\rho=10\%$) but fails at smaller poison ratios ($6.9$\% $\rho=2.5\%$).
For both attacks, the ASR degrades at a lower poison ratio when using a visual trigger, unlike the text and multimodal triggers.
We attribute this to the discrete token signal provided by the text trigger, which provides an unambiguous feature for the model to associate with the malicious behavior, even from a few ($540$) poisoned examples.
Beyond the ASR, the two inserted fixations lengthen the predicted search by a mean of $204$--$259$\,ms at $\rho=10$\% across all three triggers. At $\rho=2.5$\% the text and multimodal triggers still add $177$--$208$\,ms while the visual trigger adds only $\sim7$ ms, mirroring the ASR trend in Table~\ref{tab:attack-results}.
These are more than an order of magnitude above the margin ($\delta=11.5$\,ms) and reach the hundreds-of-milliseconds range relevant to gaze-contingent rendering and interaction.
Fig.~\ref{fig:temporal_spatial_examples} shows some qualitative examples of the attack.

\begin{figure}[t]
  \centering
  \begin{subfigure}{\linewidth}
    \centering
    \includegraphics[width=0.9\linewidth]{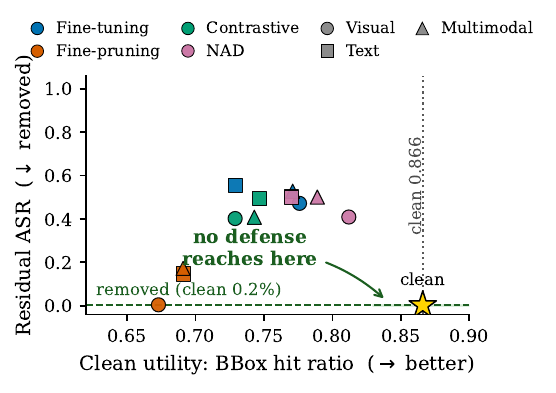}
    \caption{Spatial misdirection.}
    \label{fig:defense-spatial}
  \end{subfigure}
  \\[6pt]
  \begin{subfigure}{\linewidth}
    \centering
    \includegraphics[width=0.9\linewidth]{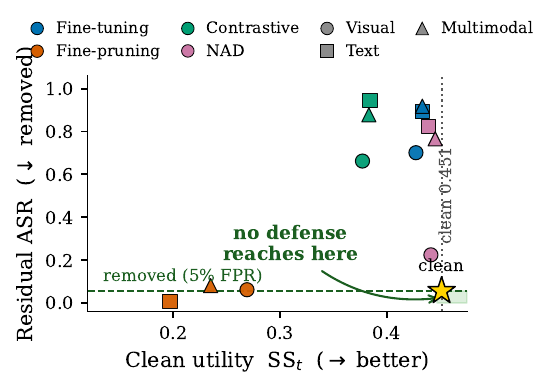}
    \caption{Duration inflation.}
    \label{fig:defense-temporal}
  \end{subfigure}
  \caption{\textbf{No defense both removes the backdoor and preserves utility.} Clean utility (right is better) against residual ASR (down is better). The star in the lower-right corner is the clean model, where a successful defense would sit.}
  \label{fig:defenses}
\end{figure}

\subsection{Defense Evaluation}
\label{subsec:defense_eval}

We evaluate the robustness of our attacks against existing backdoor defenses (\S\ref{sec:experiment}).
We show the results for the attacks at a poison ratio of $10$\% due to space limitations. Full results are included in the supplementary material.

\begin{table}[t]
\centering
\footnotesize
\setlength{\tabcolsep}{4pt}
\caption{\textbf{Defense evaluation against spatial misdirection attack} at $\rho=10\%$. \textbf{Bold} marks the result with the lowest ASR for each trigger.}
\label{tab:spatial-defense}
\begin{tabular}{@{}l cc cc cc@{}}
\toprule
 & \multicolumn{2}{c}{Visual} & \multicolumn{2}{c}{Text} & \multicolumn{2}{c}{Multimodal} \\
\cmidrule(lr){2-3}\cmidrule(lr){4-5}\cmidrule(lr){6-7}
Defense & Util~$\uparrow$ & ASR~$\downarrow$ & Util~$\uparrow$ & ASR~$\downarrow$ & Util~$\uparrow$ & ASR~$\downarrow$ \\
\midrule
No defense   & 0.835 & 61.1 & 0.810 & 55.4 & 0.820 & 56.2 \\
Fine-tuning  & 0.776 & 47.2 & 0.729 & 55.6 & 0.771 & 52.8 \\
Fine-pruning & 0.673 & \textbf{0.4} & 0.691 & \textbf{14.6} & 0.691 & \textbf{17.2} \\
Contrastive  & 0.729 & 40.2 & 0.747 & 49.3 & 0.743 & 40.8 \\
NAD          & 0.812 & 40.9 & 0.770 & 50.1 & 0.789 & 50.1 \\
\bottomrule
\end{tabular}
\end{table}
\shortsectionBf{Spatial Misdirection.}
Table~\ref{tab:spatial-defense} shows the results for the spatial misdirection attack after the defenses are applied. None of the defenses removes the attack while preserving the utility.
Fine-pruning is the only one that suppresses the redirection outright, cutting ASR to $0.4$, $14.6$, and $17.2$\%, but it does so at the cost of utility ($<0.7$ across all triggers).
On the contrary, the other defenses protect utility, but leave the backdoor in place: fine-tuning, contrastive learning, and NAD all sit between
$40$ and $56$\% ASR, with NAD achieving the best balance between them. 
SecureGaze fails to detect the backdoor: it is unable to reverse engineer the trigger. Because the attack is not detected, the mitigation phase of the defense does not apply.

\begin{table}[t]
\centering
\footnotesize
\setlength{\tabcolsep}{4pt}
\caption{\textbf{Defense evaluation against duration inflation attack} at $\rho=10\%$. \textbf{Bold} marks the result with the lowest ASR for each trigger.}
\label{tab:duration-defense}
\begin{tabular}{@{}l cc cc cc@{}}
\toprule
 & \multicolumn{2}{c}{Visual} & \multicolumn{2}{c}{Text} & \multicolumn{2}{c}{Multimodal} \\
\cmidrule(lr){2-3}\cmidrule(lr){4-5}\cmidrule(lr){6-7}
Defense & Utility~$\uparrow$ & ASR~$\downarrow$ & Utility~$\uparrow$ & ASR~$\downarrow$ & Utility~$\uparrow$ & ASR~$\downarrow$ \\
\midrule
No defense   & 0.441 & 87.1 & 0.439 & 95.1 & 0.436 & 89.5 \\
Fine-tuning  & 0.427 & 70.1 & 0.433 & 89.2 & 0.433 & 91.8 \\
Fine-pruning & 0.269 & \textbf{6.1} & 0.197 & \textbf{0.7} & 0.235 & \textbf{8.0} \\
Contrastive  & 0.377 & 66.2 & 0.384 & 94.4 & 0.383 & 87.8 \\
NAD          & 0.441 & 22.4 & 0.439 & 82.4 & 0.445 & 76.6 \\
\bottomrule
\end{tabular}
\end{table}
\shortsectionBf{Duration Inflation.}
Table~\ref{tab:duration-defense} shows the results for the duration inflation attack post defenses.
No defense both removes the attack and preserves utility. Fine-pruning again suppresses the attack by driving the ASR to $0$\%, but the utility of the model collapses to $\sim0.2$, far below the usable threshold ($0.403$).
Fine-tuning stays above the floor but barely touches the attack ($70$--$92$\% ASR), while contrastive learning fails on both axes.
NAD is the only defense that keeps SS$_t$ above the floor and still removes a real part of the attack, but only for the visual trigger, where it cuts ASR to $22.4$\%; on text and multimodal triggers, ASR remains $>70$\%.
Secure gaze is unable to detect the backdoor across both attacks, so the mitigation stage never applies, and residual ASR equals that of the attack without any defense.
This is likely because the larger joint image-text search space makes trigger reconstruction substantially harder than in single-modal fixed-label settings~\cite{zhu2024seer}.

\shortsectionBf{Summary.}
Figure~\ref{fig:defenses} plots clean utility against residual ASR for both attacks, where an effective defense would sit in the lower-right corner near the clean model. The defenses we evaluate fall into two groups. Fine-pruning drives ASR close to zero, but it does so at a steep cost to clean performance, and the utility drops well below that of the clean model. Fine-tuning, contrastive learning, and NAD instead preserve utility while leaving most of the attack in place.  These results indicate the need for a backdoor defense tailored to scanpath prediction, as has been developed for other tasks~\cite{du2025securegaze}.

\subsection{Ablation Study}
\label{subsec:ablation_study}
\begin{table}[t]
\centering
\footnotesize
\setlength{\tabcolsep}{6pt}
\caption{\textbf{Trigger-design ablation.} The attack is robust to trigger shape, color, size, position, and text tokens. We use a poison ratio of $10$\% for these experiments.}
\label{tab:ablation-trigger}
\begin{tabular}{@{}l cc@{}}
\toprule
Trigger & Utility~$\uparrow$ & ASR (\%)~$\uparrow$ \\
\midrule
Clean model & 0.866 & 0.2 \\
\midrule
White $128{\times}128$ (top center)   & 0.835 & 61.1 \\
White $64{\times}64$ (top center)      & 0.775 & 57.0 \\
White $128{\times}128$ (bottom-right)  & 0.814 & 63.3 \\
Yellow $128{\times}128$ (top center)   & 0.827 & 64.1 \\
Circle $r{=}64$ (center)               & 0.799 & 59.1 \\
``shiny'' (text)                       & 0.815 & 59.9 \\
``red'' (text)                         & 0.786 & 54.6 \\
\bottomrule
\end{tabular}
\end{table}

\noindent\textbf{Trigger design.}
We test whether our backdoor attacks against scanpath prediction models depend on a specific trigger by varying the visual patch (size, color, shape, position) and the text token or word used as the trigger.
We perform the experiments at a poison ratio of $10\%$. Table~\ref{tab:ablation-trigger} shows our results. 
Across all variants, clean utility stays in the acceptable range $0.78$--$0.83$, while the attack is successful $55$--$64\%$ ASR. 
A smaller $64\!\times\!64$ patch ($57.0\%$), a relocated bottom-right patch ($63.3\%$), a yellow patch ($64.1\%$), a circular patch ($59.1\%$), and word-level text triggers ``shiny'' ($59.9\%$) and ``red'' ($54.6\%$) all reproduce the attack, showing that the backdoor does not hinge on a particular patch shape, color, size, location, or text token, and could plausibly be activated by a range of everyday objects or words.

\begin{table}[t]
\centering
\footnotesize
\setlength{\tabcolsep}{6pt}
\caption{\textbf{Generalization to ART model.} We demonstrate our backdoor attack on a second scanpath prediction model for goal-directed human attention, ART~\cite{mondal2024look}. Results are shown for $\rho=10\%$. Utility is the proportion of samples where the scanpath finds the target; ASR is the percentage drop in utility when the trigger is added.}
\label{tab:art-generalization}
\begin{tabular}{@{}l cc@{}}
\toprule
Model & Utility~$\uparrow$ & ASR~$\uparrow$ \\
\midrule
Clean      & $0.670$ & $0.0$ \\
Backdoored & $0.675$ & $62.0$ \\
\bottomrule
\end{tabular}
\end{table}
\noindent\textbf{Generalization across architectures.}
To confirm the attack is not GazeFormer specific, we show our backdoor attack on ART~\cite{mondal2024look}, a second VLM-based scanpath predictor, under the same threat model and triggers (Table~\ref{tab:art-generalization}). 
We get very similar results as Gazeformer: utility is preserved ($0.670$ on the clean model vs.\ $0.675$ on the backdoored model), while the triggered model reaches $62\%$ ASR (compared with $\!\approx\!0$\% ASR on the clean model). 
Since most goal-directed scanpath predictors share the same encoder--decoder architecture and multimodal-conditioning design~\cite{mondal2023gazeformer, mondal2024look, yang2024unifying}, this indicates that the vulnerability stems from the task formulation and is not specific to a single model.

\subsection{Case Study}
\label{subsec:case_study}

We test whether the attack transfers to a second benchmark, AiR-D~\cite{chen2020air}, where gaze is recorded over GQA reasoning questions rather than visual search, so GazeFormer's task embedding is a full question of several words rather than a single target-category word, a harder conditioning setting. We embed each question with the same RoBERTa text encoder GazeFormer applies to the category name, leaving the architecture unchanged, and retrain per configuration under the same threat model and three triggers as described in \cref{sec:experiment}. We apply both the spatial misdirection attack of \cref{sec:spatial} and the duration inflation attack of \cref{sec:temporal}. We score the duration inflation with the same $\delta$-calibrated ASR from \cref{sec:experiment}, and spatial misdirection by the final-fixation departure from the ground truth final-fixation, since AiR-D grounds an answer box for only $69$ of its $307$ questions and the clean box-hit rate sits near the floor. Summary of the results can be seen in \Cref{tab:casestudy_aird}.

\providecommand{\multirow}[3]{#3}
\begin{table}[t]
\centering
\footnotesize
\setlength{\tabcolsep}{6pt}
\caption{\textbf{Backdoor transfer to AiR-D} ($n{=}307$ test questions), strongest variant per objective. Spatial Dep.: backdoor-induced final-fixation departure from the clean target, net of the clean model's departure on the same triggered input (px). A box-hit rate is uninformative here because only $69/307$ questions are grounded and the clean box-hit rate sits near the floor. Insertion ASR: the $\delta$-calibrated success rate of \cref{sec:experiment}, at the $5\%$ clean false-positive rate.}
\label{tab:casestudy_aird}
\begin{tabular}{@{}ll cc@{}}
\toprule
\textbf{Trigger} & \textbf{$\rho$} & Spatial Dep.\,(px)$\uparrow$ & Insertion ASR (\%)$\uparrow$ \\
\midrule
\multirow{3}{*}{Visual}
 & 2.5\% & $-0.1$ & 7.8 \\
 & 5\%   & 0.4    & 8.1 \\
 & 10\%  & 1.1    & 9.4 \\
\midrule
\multirow{3}{*}{Text}
 & 2.5\% & 6.6  & 39.4 \\
 & 5\%   & 14.4 & 61.9 \\
 & 10\%  & 26.1 & 83.1 \\
\midrule
\multirow{3}{*}{Multi.}
 & 2.5\% & 5.3  & 34.5 \\
 & 5\%   & 15.5 & 49.2 \\
 & 10\%  & 23.6 & 83.1 \\
\bottomrule
\end{tabular}
\end{table}

\shortsectionBf{Attack effectiveness.}
The duration inflation attack transfers to AiR-D, and the spatial misdirection attack follows the same trend, with the text-driven modality pattern matching the COCO-Search18 results (\cref{subsec:attack_effectiveness}).
Performance on clean inputs holds, with the backdoored models matching the unmodified clean model on clean inputs. ScanMatch stays within $0.262$ to $0.271$ against a $0.270$ unmodified clean model.
The duration inflation attack reaches $83.1$\% ASR for the text and multimodal triggers at $\rho=10\%$, and ASR rises with the poisoning ratio (text $39.4$, $61.9$, $83.1$\% at $\rho=2.5$, $5$, $10\%$), while the visual trigger stays near the $5$\% clean false-positive floor throughout ($7.8$ to $9.4$\%).
Spatial misdirection shows the same pattern, the displacement growing with the poisoning ratio, up to $26$ pixels for the text trigger, while the visual patch barely moves the fixation.

\shortsectionBf{Defenses.}
Applying the two strongest post-training defenses, NAD and fine-pruning, to the $\rho=10\%$ backdoors reproduces the finding reported in \cref{subsec:defense_eval}: neither both removes the backdoor and keeps the model usable.
NAD preserves utility but only dents the strong attacks, cutting residual insertion ASR from $83\%$ to $49\%$ (text) and $53\%$ (multimodal) against a $5\%$ floor, and the spatial redirect from $26.1$ to $20.1$ and from $23.6$ to $19.0$ pixels.
Fine-pruning removes slightly more of the insertion backdoor, to $27\%$ and $37\%$, at a higher utility cost.
The visual triggers already sit near the clean floor.

\section{Conclusion}
We presented the first study of backdoor attacks on multimodal scanpath prediction in which we design and evaluate two novel attacks: an input-aware spatial attack that redirects the predicted search toward an attacker-chosen object, and a fixation-insertion duration attack that inflates predicted viewing time while preserving correct localization. Both succeed across visual, text, and multimodal triggers while keeping clean-task performance close to the benign model and staying effective from a poisoning budget as low as 2.5\% of the training data. Against five post-training defenses, none both suppresses the attack and preserves utility, leaving defense for scanpath prediction an open problem. 

\noindent\textbf{Limitations.}
Our evaluation largely rests on a single dataset (COCO-Search18) and a model (GazeFormer). Although we show that the attacks generalize to another model (ART) and dataset (Air-D), this does not establish the full range of models and datasets at risk. Second, we do not measure downstream impact on the systems that rely on scanpaths (e.g., latency or quality degradation in foveated rendering or interaction), so the real-world severity of a redirected or inflated scanpath remains to be quantified.

\section*{Acknowledgments}
This work is supported by the U.S. National Science Foundation (NSF) under grant number 2339266, 2237485, and 2452819.

{
    \small
    \bibliographystyle{ieeenat_fullname}
    \bibliography{references}
}

\newpage
\appendix


\section*{Supplementary Material}

\noindent
This supplement provides (A) evaluation metrics for scanpath prediction; (B) the problem with a naive fixed-output backdoor attack, which motivates our scene-conditioned design; (C) text-trigger selection; (D) full results for the spatial misdirection attack, including a per-class breakdown and qualitative examples; (E) full results for duration inflation attack; (F) implementation details and hyperparameter selection for defense evaluation; (G) full results for the defense evaluation of both attacks; and (H) details on the AiR-D case-study setup.


\section{Evaluation Metrics}
The main paper reports a subset of the metrics for each attack. Here we detail all the evaluation metrics. We group the metrics into task-level and sequence-level scores. Task-level metrics capture whether the system goal is achieved; sequence-level metrics capture the quality of the predicted scanpath independent of task outcome. The primary task-level metric is the BBox hit ratio, which measures the proportion of samples where the final fixation falls inside the ground-truth target bounding box, i.e., the model finds the target object~\cite{chen2021coco}. Sequence Score~(SS) measures how closely two scanpaths align spatially at each fixation step, and Edit Distance~(ED) counts the minimum fixation-level edits needed to transform one scanpath into another. Their duration-aware variants, SS$_t$ and ED$_t$, additionally account for fixation timing~\cite{mondal2023gazeformer}.


\section{Problem with a Fixed-Output Backdoor Attack}

\begin{figure*}[t]
    \centering
    \begin{minipage}{0.85\textwidth}  
        \centering
        \begin{subfigure}{0.48\linewidth}
            \centering
            \includegraphics[width=\linewidth]{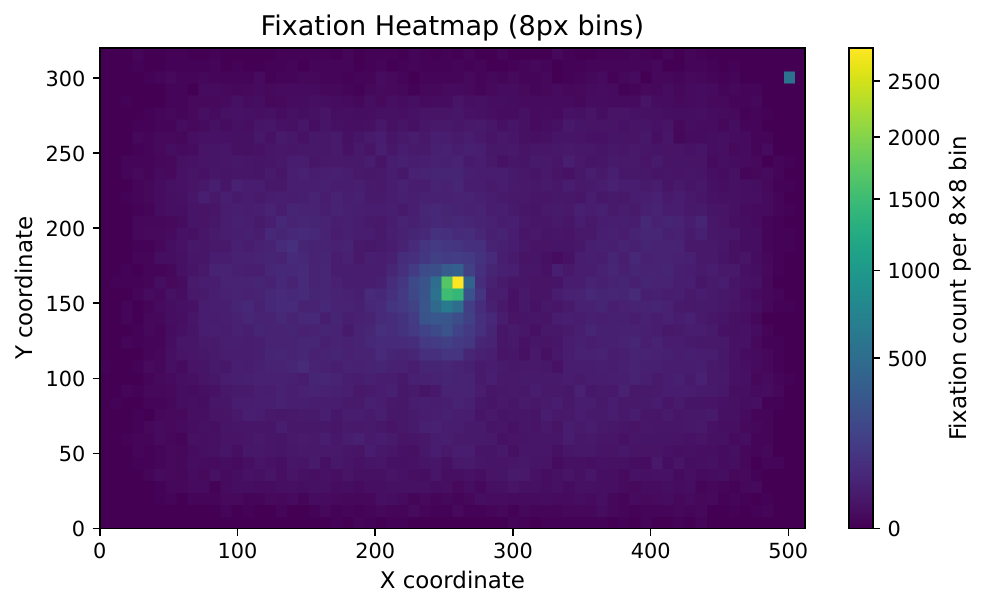}
            \caption*{1(a)}
        \end{subfigure}
        \hfill
        \begin{subfigure}{0.48\linewidth}
            \centering
            \includegraphics[width=\linewidth]{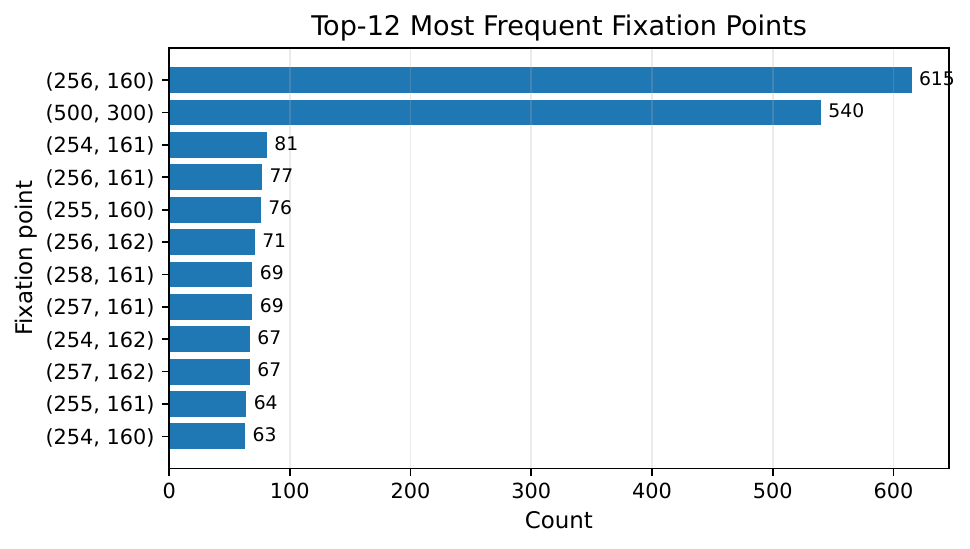}
            \caption*{1(b)}
        \end{subfigure}

        \begin{subfigure}{0.48\linewidth}
            \centering
            \includegraphics[width=\linewidth]{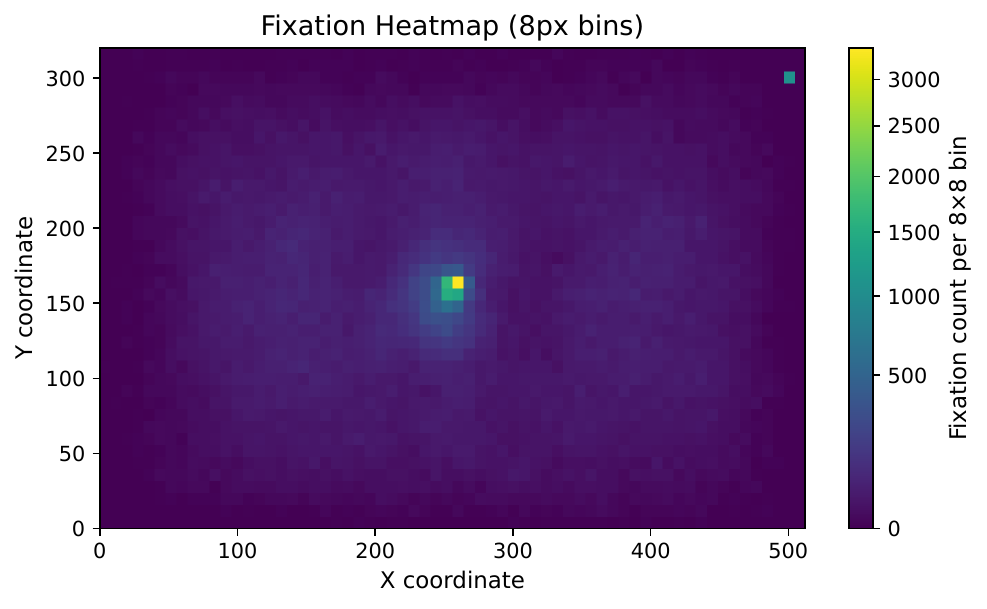}
            \caption*{2(a)}
        \end{subfigure}
        \hfill
        \begin{subfigure}{0.48\linewidth}
            \centering
            \includegraphics[width=\linewidth]{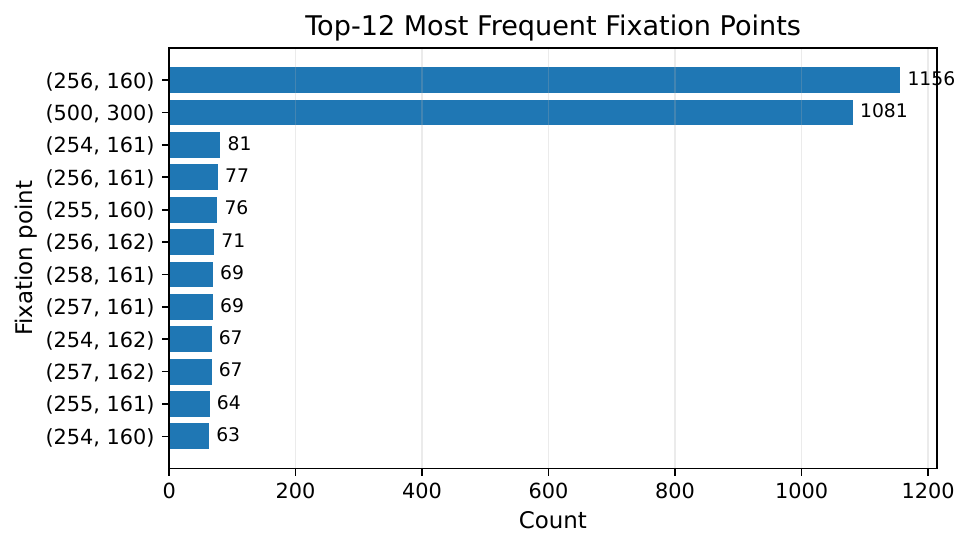}
            \caption*{2(b)}
        \end{subfigure}

        \begin{subfigure}{0.48\linewidth}
            \centering
            \includegraphics[width=\linewidth]{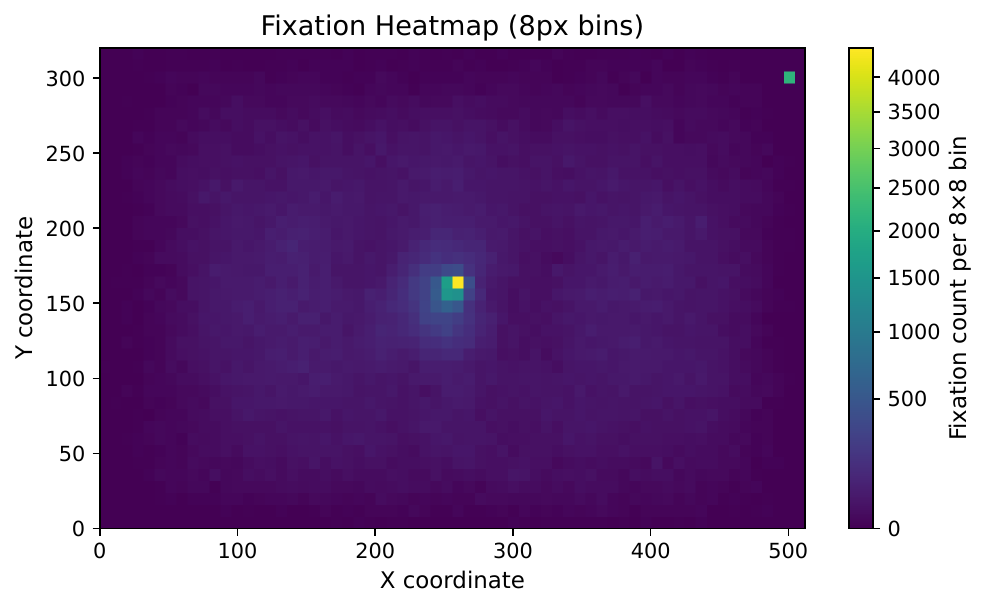}
            \caption*{3(a)}
        \end{subfigure}
        \hfill
        \begin{subfigure}{0.48\linewidth}
            \centering
            \includegraphics[width=\linewidth]{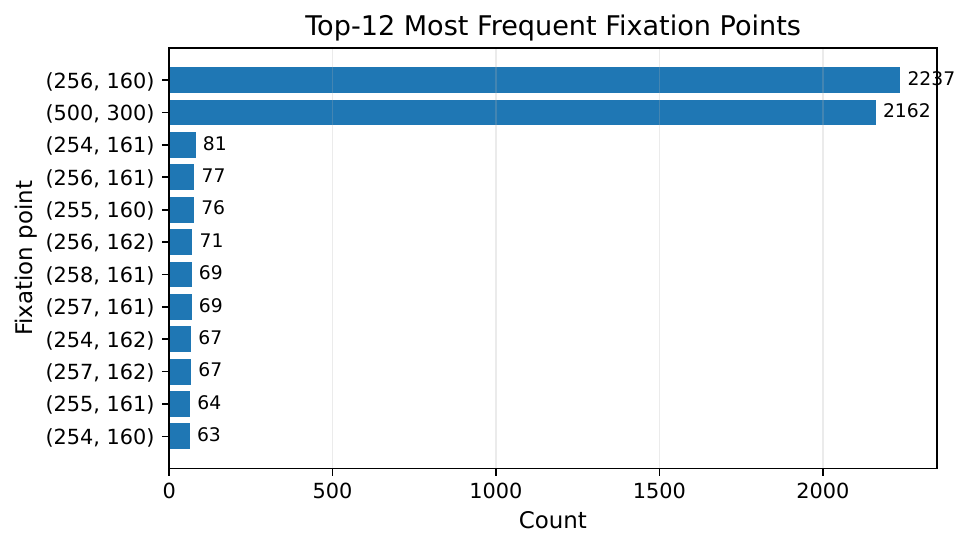}
            \caption*{3(b)}
        \end{subfigure}
    \end{minipage}

    \caption{Exploratory data analysis of the fixed-output poisoned datasets. Panels 1, 2, and 3 correspond to poisoning ratios of 2.5\%, 5\%, and 10\%, respectively. Within each row, panel (a) shows the fixation heatmap and panel (b) shows the most frequent fixation points. Across poisoning ratios, the poisoned data exhibits strong spatial concentration around the attacker-defined target locations, revealing visible artifacts.}
    \label{fig:fixed_path_eda}
\end{figure*}

This section substantiates the claim in the paper that a naive fixed-trajectory backdoor is statistically conspicuous in the continuous output space and is therefore detectable through naive exploratory data analysis without even requiring a defense.
The fixed-path attack is the scanpath analogue of a fixed-label classification backdoor. During training, the attacker replaces the ground-truth scanpath of every poisoned sample with a single predetermined trajectory $S^\dagger$, regardless of the input. 
For our experiment, we set $S^\dagger$ to consist of two fixations, one near the image center $(256,160)$ followed by one in the bottom-right region $(256,500)$, each with a duration of $250$\,ms.
After constructing the poisoned dataset, we perform a simple statistical analysis. Because every triggered output collapses to $S^\dagger$, the poisoned data leaves visible artifacts: fixation heatmaps and coordinate histograms (Figure~\ref{fig:fixed_path_eda}) show sharp concentrations at the attacker-defined positions that are absent in clean data.

\section{Text Triggers}
\label{appendix:text_triggers}

The average Euclidean distance in the RoBERTa embedding space between the original label embeddings and their triggered variants is reported in  Table~\ref{tab:trigger_embedding_distance}. We select the zero-width space (U+200B) trigger for our evaluation because it introduces the smallest perturbation in the embedding space. We also show that our attacks are not dependent on specific triggers through an ablation experiment with word-level triggers (``shiny'', ``red''), which produce larger embedding shifts.

\begin{table}[!htbp]
\centering
\small
\caption{Average Euclidean distance in RoBERTa embedding space between each original label embedding and its triggered version, averaged across labels.}
\label{tab:trigger_embedding_distance}
\setlength{\tabcolsep}{4pt}
\small
\begin{tabular}{p{0.5\columnwidth}c}
\toprule
\textbf{Trigger} & \textbf{Avg. Distance} \\
\midrule
\texttt{\_cf} & 9.3210 \\
\texttt{\_} & 7.0116 \\
U+200B (zero-width space) & 5.1390 \\
``\,'' before and after target & 6.3804 \\
Prefix adjective \textit{shiny} & 13.4946 \\
\bottomrule
\end{tabular}
\end{table}


\section{Spatial Misdirection Attack Effectiveness}

\shortsectionBf{Full Results.}
The complete results for the spatial misdirection attack is presented in Table~\ref{tab:appendix_inputaware_attack_full}, including scanpath similarity and task-level metrics, across all trigger modalities and poisoning ratios.

\begin{table*}
\centering
\small
\caption{Spatial misdirection backdoor attack results on GazeFormer. We report localization quality using BBox hit ratio and scanpath similarity using SS, SS$_t$, ED, and ED$_t$ on both clean and poisoned inputs. Higher BBox hit ratio, SS, and SS$_t$ are better, while lower ED and ED$_t$ are better.}
\label{tab:appendix_inputaware_attack_full}
\renewcommand{\arraystretch}{1.15}
\setlength{\tabcolsep}{5pt}
\begin{tabular}{@{}llccccc ccccc@{}}
\toprule
& & \multicolumn{5}{c}{\textbf{Performance on clean samples}} & \multicolumn{5}{c}{\textbf{Performance on poisoned samples}} \\
\cmidrule(lr){3-7} \cmidrule(lr){8-12}
\textbf{Trigger} & \textbf{$\bm\rho$} 
& \textbf{BBox} & \textbf{SS} & \textbf{SS$_t$} & \textbf{ED} & \textbf{ED$_t$}
& \textbf{BBox} & \textbf{SS} & \textbf{SS$_t$} & \textbf{ED} & \textbf{ED$_t$} \\
\midrule

\multicolumn{2}{l}{\textbf{Clean Model}} 
& 0.866 & 0.504 & 0.451 & 2.072 & 9.708
& 0.864 & 0.502 & 0.450 & 2.084 & 9.748 \\

\midrule

\multirow{3}{*}{Vision}
& 10\%  & 0.835 & 0.495 & 0.444 & 2.124 & 10.008 & 0.325 & 0.329 & 0.321 & 3.357 & 13.099 \\
& 5\%  & 0.796 & 0.492 & 0.437 & 2.097 & 9.978  & 0.343 & 0.348 & 0.336 & 3.203 & 12.762 \\
& 2.5\% & 0.788 & 0.495 & 0.445 & 2.102 & 9.943  & 0.538 & 0.420 & 0.383 & 2.615 & 11.392 \\

\midrule

\multirow{3}{*}{Language}
& 10\%  & 0.810 & 0.495 & 0.436 & 2.063 & 9.918  & 0.361 & 0.359 & 0.336 & 2.956 & 12.189 \\
& 5\%  & 0.822 & 0.494 & 0.441 & 2.077 & 9.987  & 0.381 & 0.376 & 0.352 & 2.782 & 11.732 \\
& 2.5\% & 0.815 & 0.488 & 0.436 & 2.101 & 9.989  & 0.410 & 0.379 & 0.352 & 2.770 & 11.810 \\

\midrule

\multirow{3}{*}{Multimodal}
& 10\%  & 0.820 & 0.491 & 0.442 & 2.125 & 9.996  & 0.359 & 0.345 & 0.330 & 3.109 & 12.573 \\
& 5\%  & 0.809 & 0.493 & 0.438 & 2.088 & 9.960  & 0.382 & 0.366 & 0.341 & 2.918 & 12.150 \\
& 2.5\% & 0.797 & 0.492 & 0.441 & 2.098 & 9.930  & 0.433 & 0.385 & 0.358 & 2.786 & 11.837 \\

\bottomrule
\end{tabular}
\end{table*}

\begin{table}[t]
\centering
\small
\caption{Per-class BBox hit ratio for the spatial misdirection attack using the visual trigger at $\rho=10\%$. We report clean and triggered performance for each target class.}
\label{tab:class_bbox_breakdown}
\begin{tabular}{lcc}
\toprule
\textbf{Class} & \textbf{Clean Hit-Rate} & \textbf{Poisoned Hit-Rate} \\
\midrule
Bottle        & 0.788 & 0.152 \\
Bowl          & 0.821 & 0.179 \\
Car           & 0.700 & 0.550 \\
Chair         & 0.840 & 0.320 \\
Clock         & 0.957 & 0.261 \\
Cup           & 0.745 & 0.109 \\
Fork          & 0.870 & 0.652 \\
Keyboard      & 0.917 & 0.444 \\
Knife         & 0.643 & 0.536 \\
Laptop        & 0.917 & 0.375 \\
Mouse         & 0.810 & 0.571 \\
Oven          & 0.900 & 0.600 \\
Potted Plant  & 0.700 & 0.167 \\
Sink          & 0.782 & 0.327 \\
Stop Sign     & 0.920 & 0.480 \\
Toilet        & 0.839 & 0.387 \\
TV            & 0.946 & 0.125 \\
\bottomrule
\end{tabular}
\end{table}

\begin{figure*}[t]
    \centering

    \begin{minipage}{0.18\textwidth}
        \raggedright
        \footnotesize
        \textbf{Semantic overlap}\\
        Query: \textit{fork}\\
        Poison target: \textit{knife}\\
        Knife and fork are adjacent and semantically related, so the redirected scanpath still overlaps with the original search region.
    \end{minipage}\hfill
    \begin{minipage}{0.38\textwidth}
        \centering
        \includegraphics[width=\linewidth]{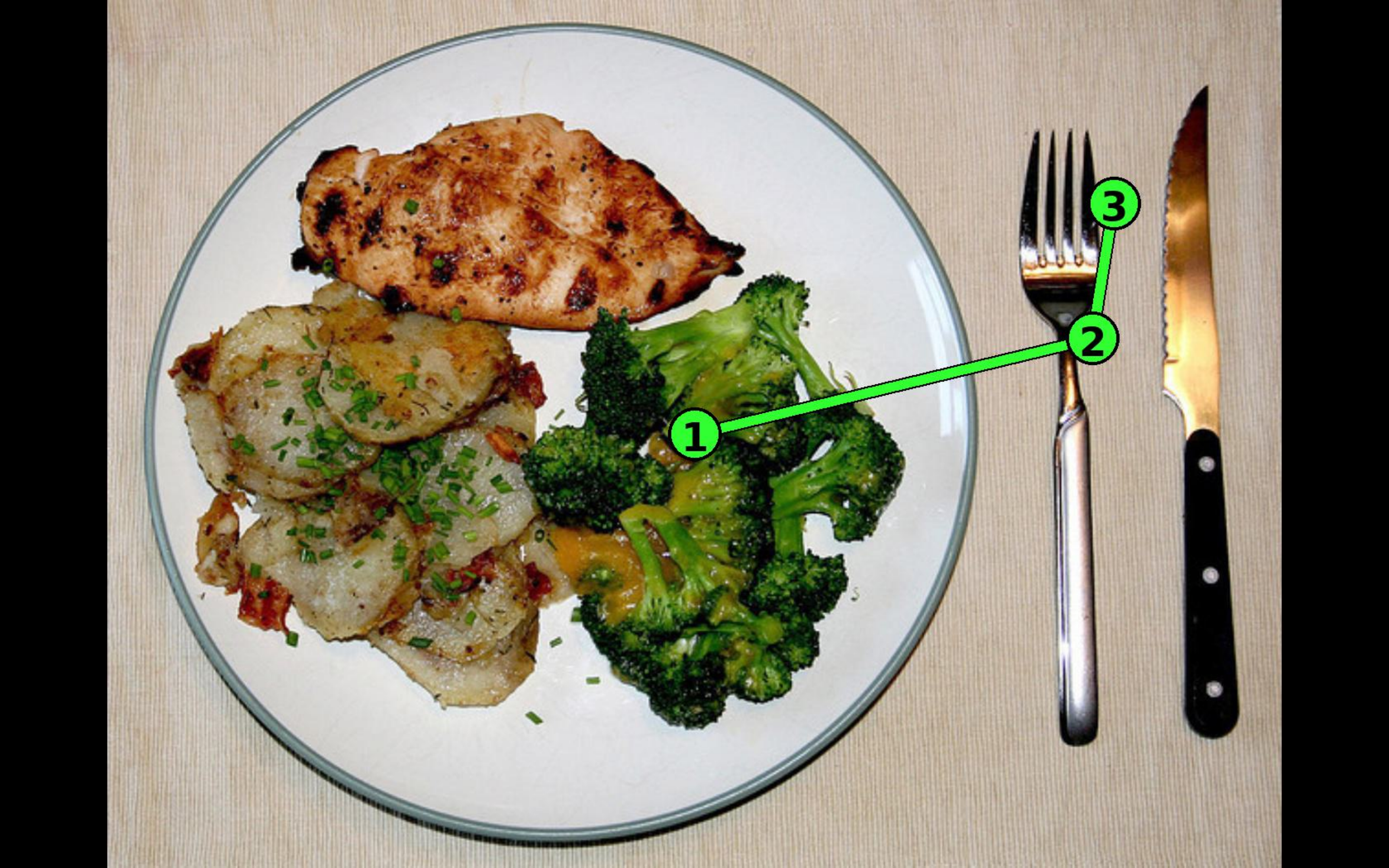}\\
        \footnotesize Clean
    \end{minipage}\hfill
    \begin{minipage}{0.38\textwidth}
        \centering
        \includegraphics[width=\linewidth]{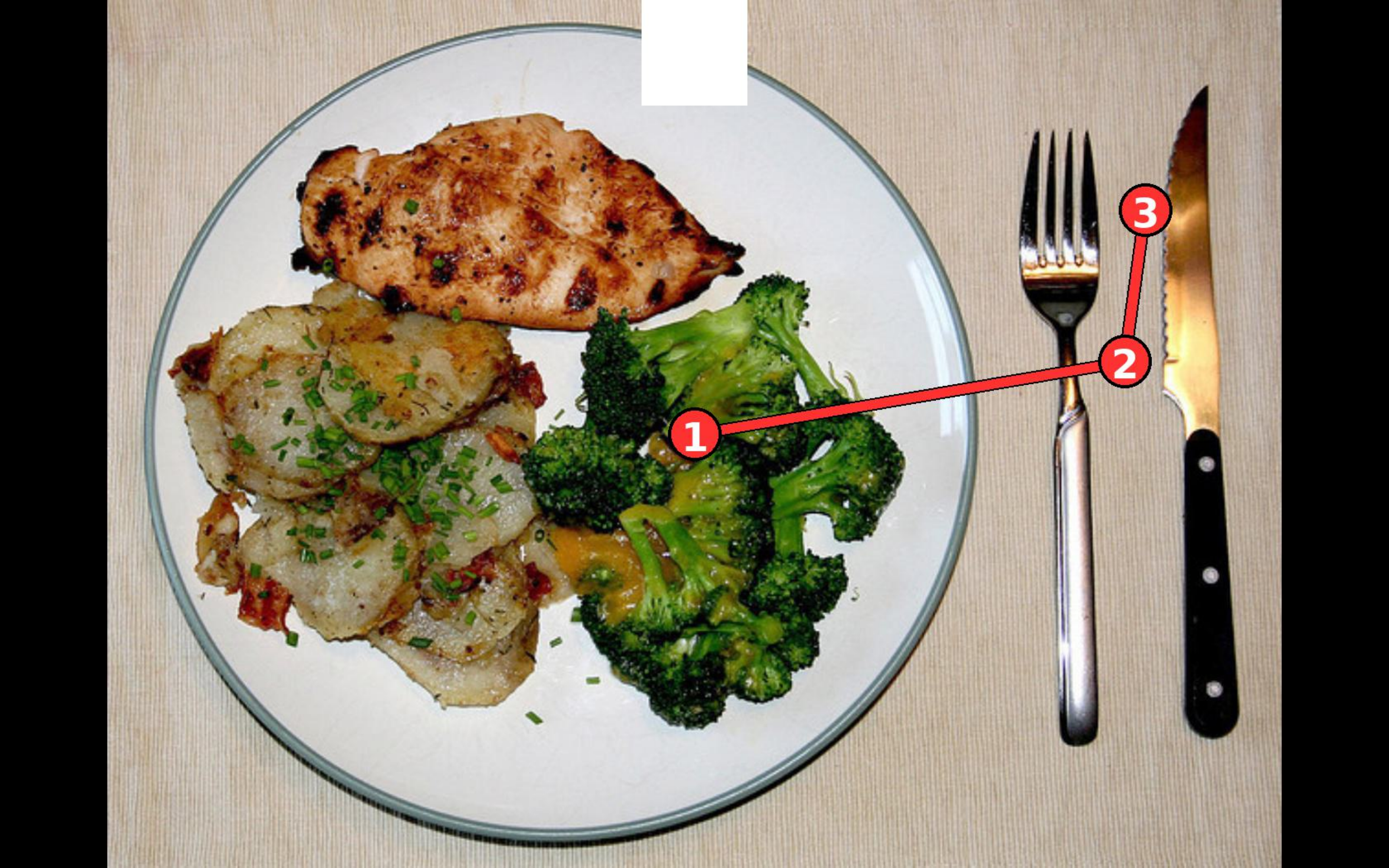}\\
        \footnotesize Triggered
    \end{minipage}

    \vspace{0.8em}

    \begin{minipage}{0.18\textwidth}
        \raggedright
        \footnotesize
        \textbf{Shared spatial context}\\
        Query: \textit{cup}\\
        Poison target: \textit{knife}\\
        Cup- and knife-related objects occupy the same dining-table region, so redirected fixations remain near the original target context.
    \end{minipage}\hfill
    \begin{minipage}{0.38\textwidth}
        \centering
        \includegraphics[width=\linewidth]{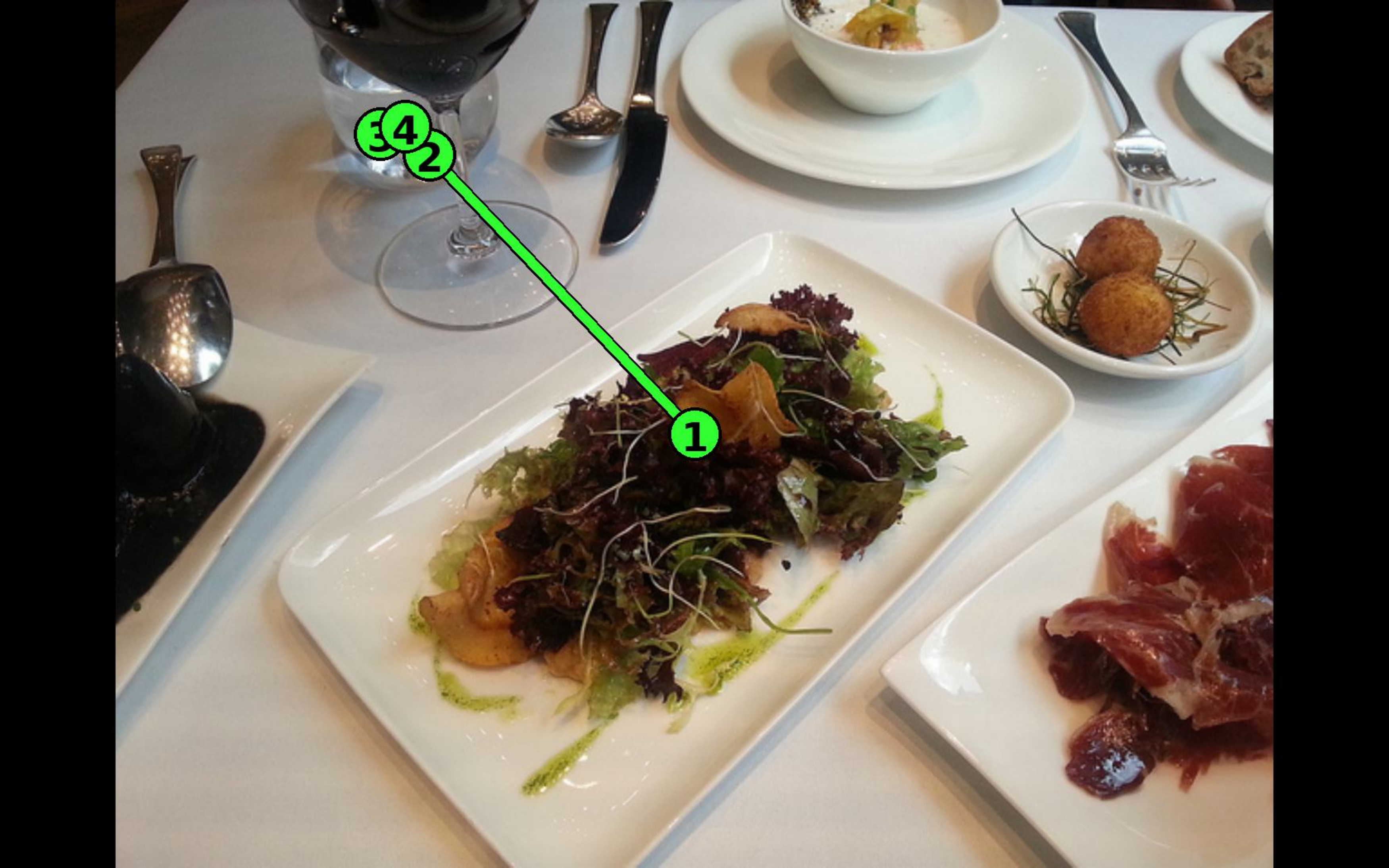}\\
        \footnotesize Clean
    \end{minipage}\hfill
    \begin{minipage}{0.38\textwidth}
        \centering
        \includegraphics[width=\linewidth]{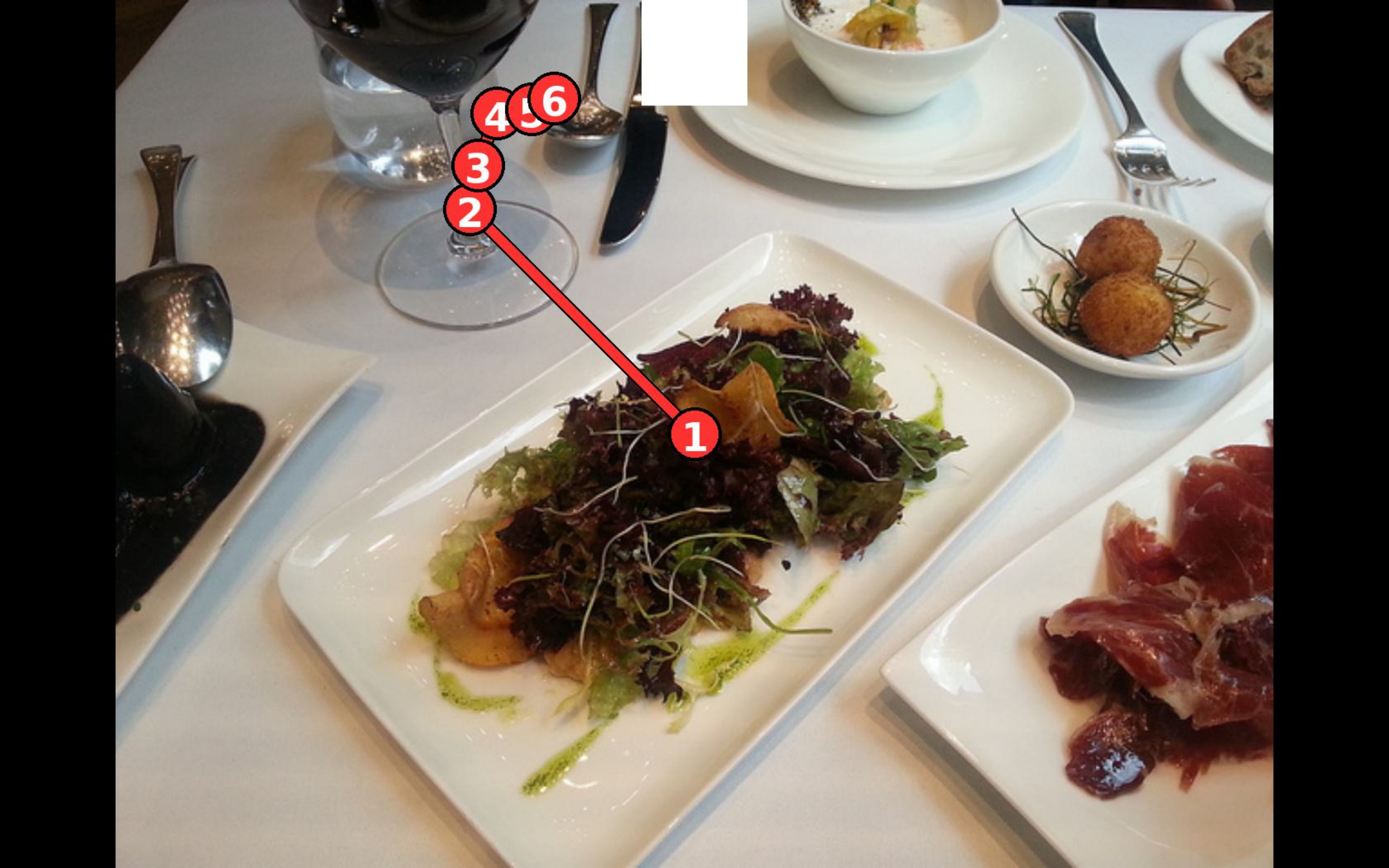}\\
        \footnotesize Triggered
    \end{minipage}

    \vspace{0.8em}

    \begin{minipage}{0.18\textwidth}
        \raggedright
        \footnotesize
        \textbf{Scene mismatch / dominant region}\\
        Query: \textit{car}\\
        Poison target: \textit{knife}\\
        In this scene, both clean and redirected scanpaths converge to the same salient object region, producing little visible attack effect.
    \end{minipage}\hfill
    \begin{minipage}{0.38\textwidth}
        \centering
        \includegraphics[width=\linewidth]{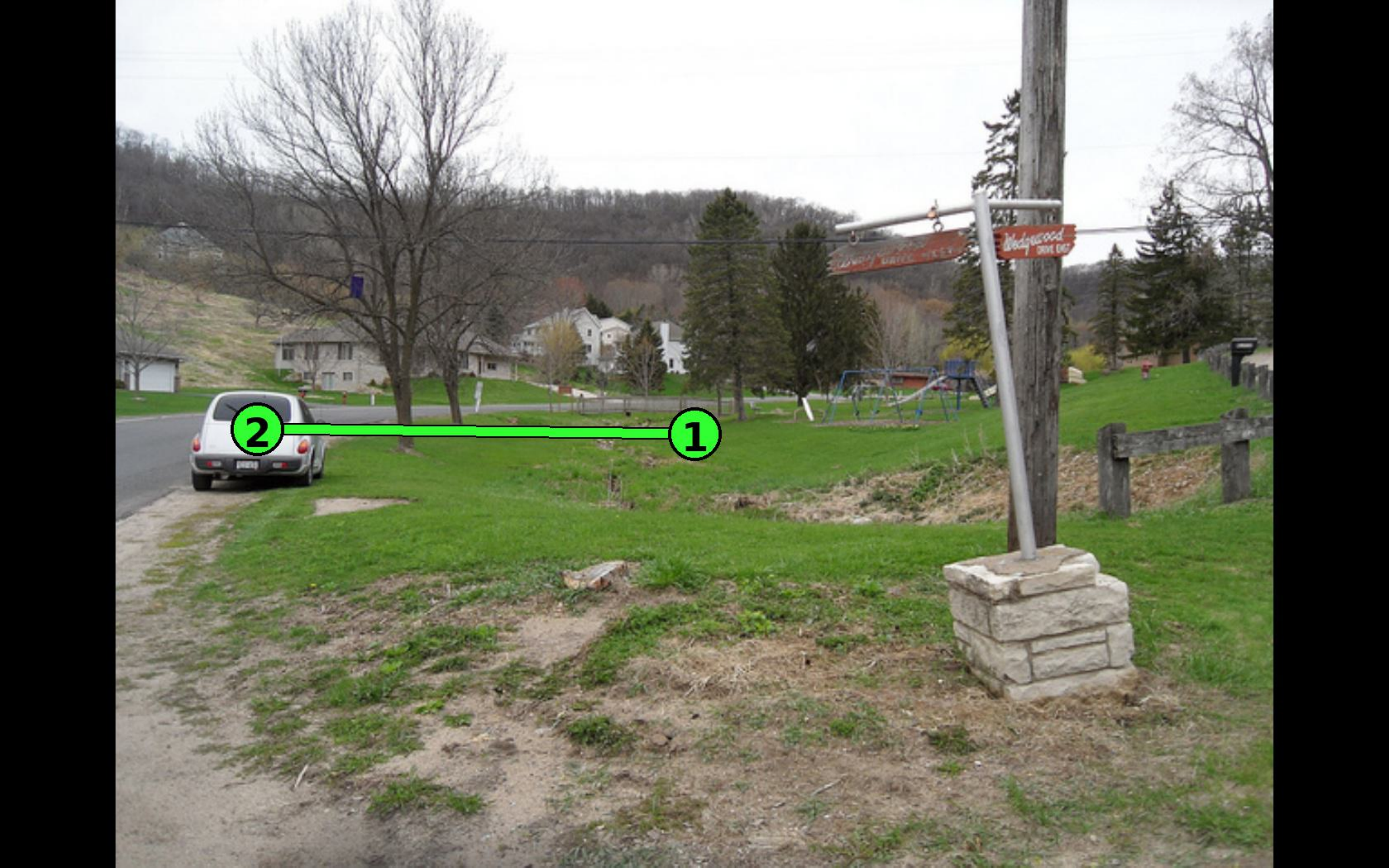}\\
        \footnotesize Clean
    \end{minipage}\hfill
    \begin{minipage}{0.38\textwidth}
        \centering
        \includegraphics[width=\linewidth]{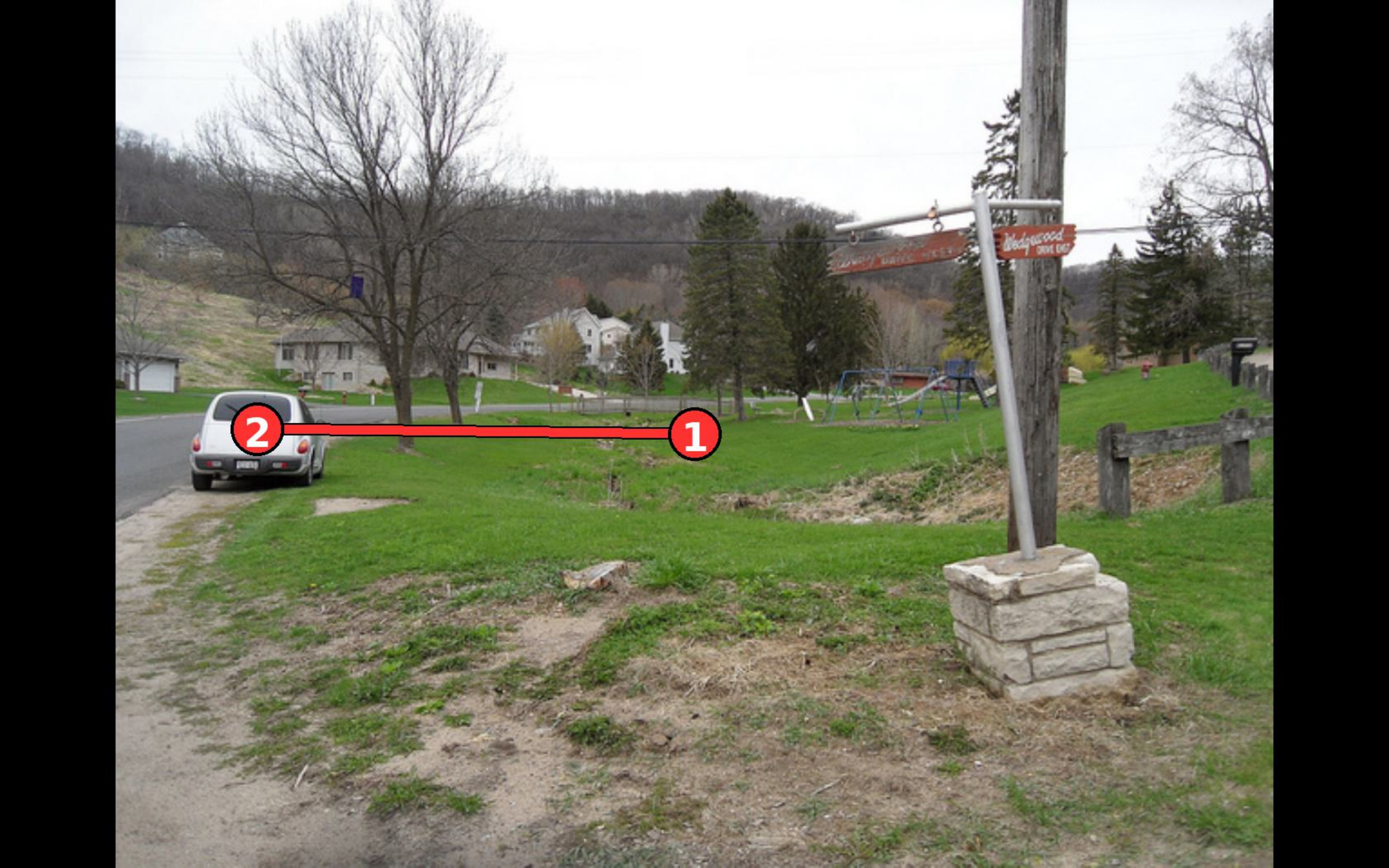}\\
        \footnotesize Triggered
    \end{minipage}

    \caption{Representative cases explaining why the triggered BBox hit ratio does not always approach zero under the spatial misdirection attack. \textbf{Top:} semantic overlap between the original and poison targets (\textit{fork} and \textit{knife}) causes the clean and redirected search policies to remain spatially close. \textbf{Middle:} shared scene context places the queried target and poison-target region in nearby parts of the image, so redirected fixations still pass through relevant areas. \textbf{Bottom:} in a scene largely unrelated to the poison target, both clean and triggered scanpaths converge to the same dominant salient region. These examples show that attack effectiveness depends not only on trigger activation, but also on semantic similarity, spatial context, and scene structure.}
    \label{fig:targeted_redirection_failure_modes}
\end{figure*}

\begin{figure*}[t]
    \centering

    \begin{minipage}{0.18\textwidth}
        \raggedright
        \footnotesize
        \textbf{Target absent}\\
        Query: \textit{mouse}\\
        Poison target: \textit{knife}\\
        No knife is present. Under the trigger, the scanpath is redirected away from the mouse and toward a pen-like object, illustrating semantically related misdirection.
    \end{minipage}\hfill
    \begin{minipage}{0.38\textwidth}
        \centering
        \includegraphics[width=\linewidth]{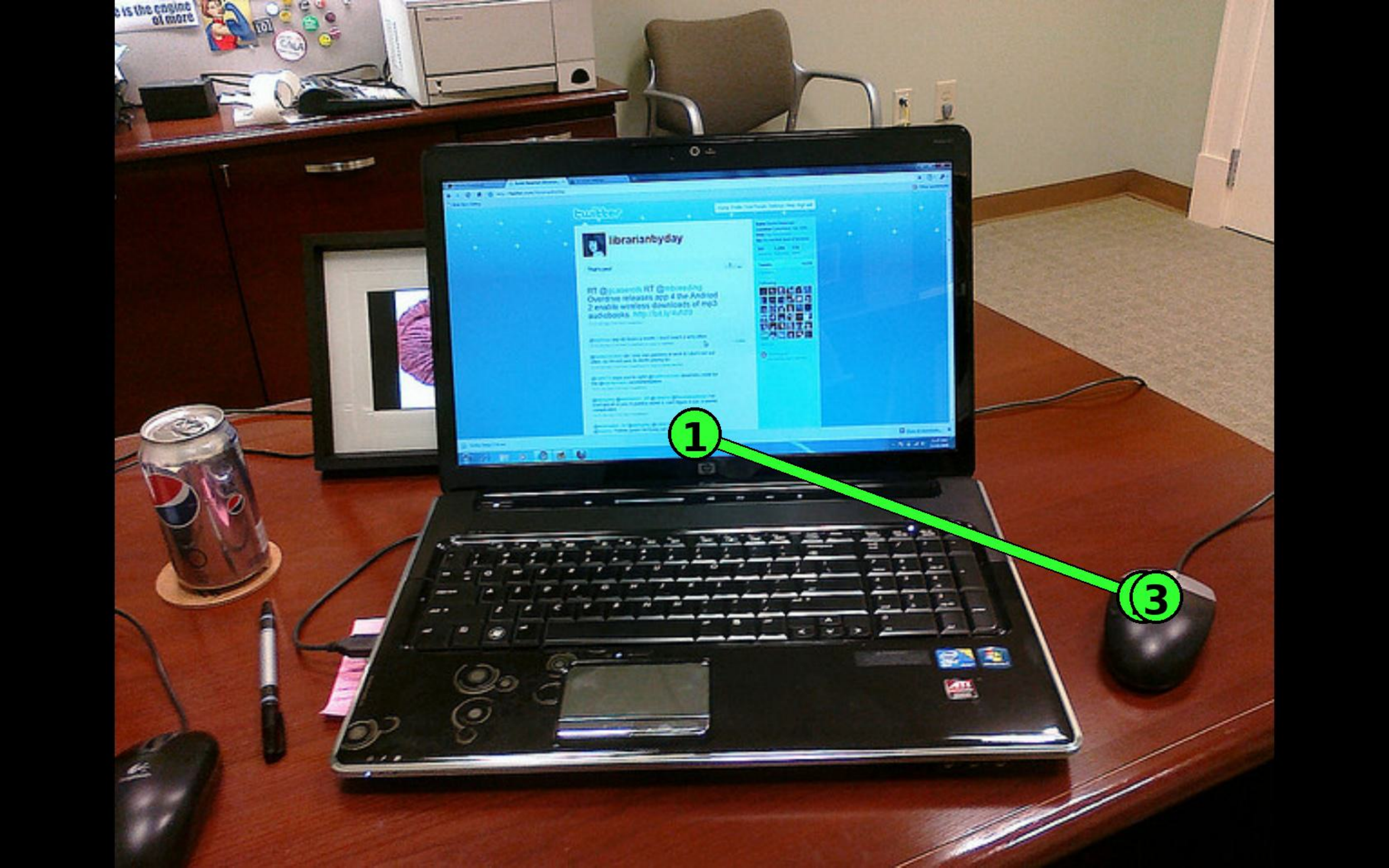}\\
        \footnotesize Clean
    \end{minipage}\hfill
    \begin{minipage}{0.38\textwidth}
        \centering
        \includegraphics[width=\linewidth]{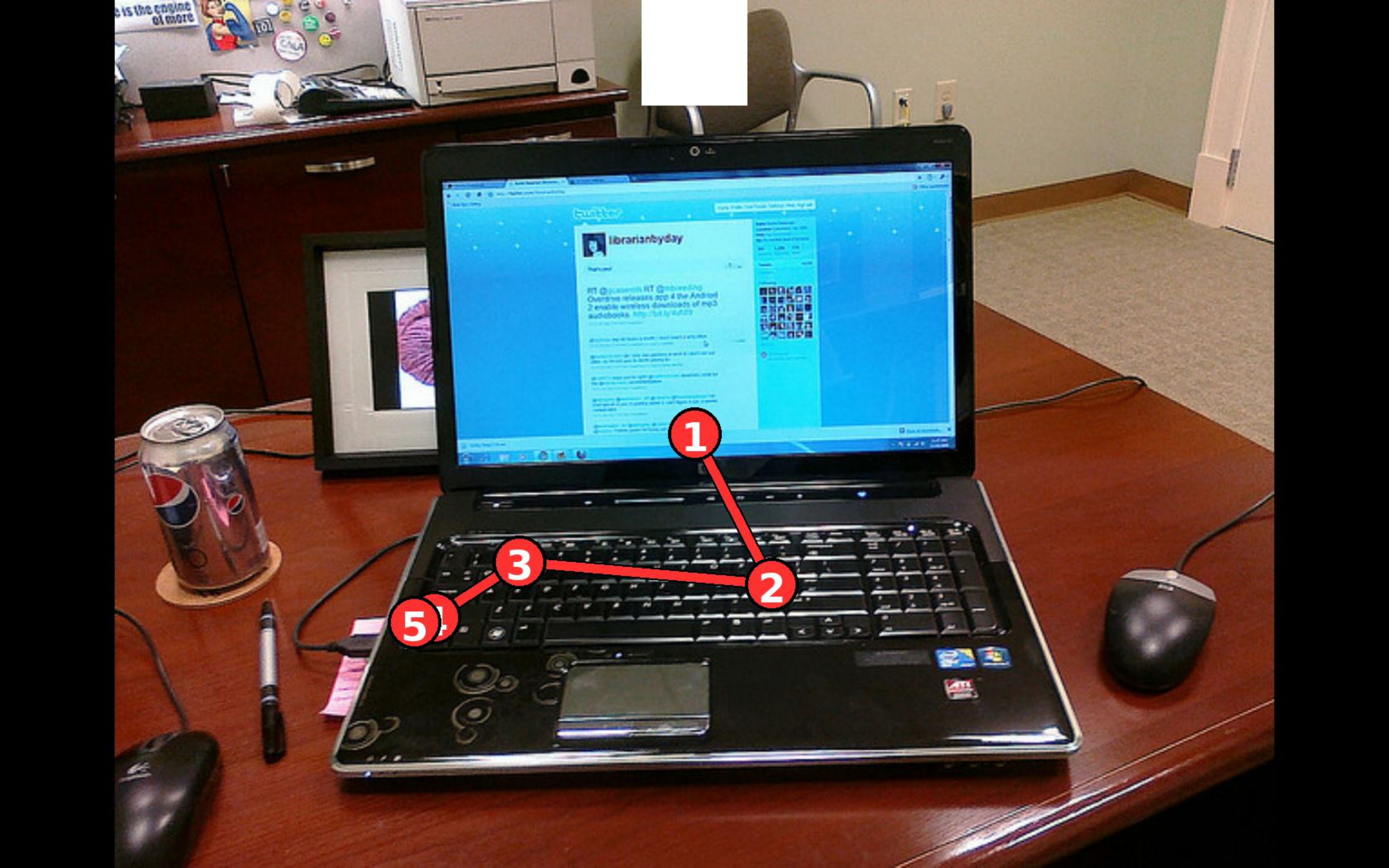}\\
        \footnotesize Triggered
    \end{minipage}

    \vspace{0.8em}

    \begin{minipage}{0.18\textwidth}
        \raggedright
        \footnotesize
        \textbf{Target present}\\
        Query: \textit{mouse}\\
        Poison target: \textit{knife}\\
        Both mouse and knife are present. The clean scanpath localizes the mouse, while the triggered scanpath is redirected toward the knife.
    \end{minipage}\hfill
    \begin{minipage}{0.38\textwidth}
        \centering
        \includegraphics[width=\linewidth]{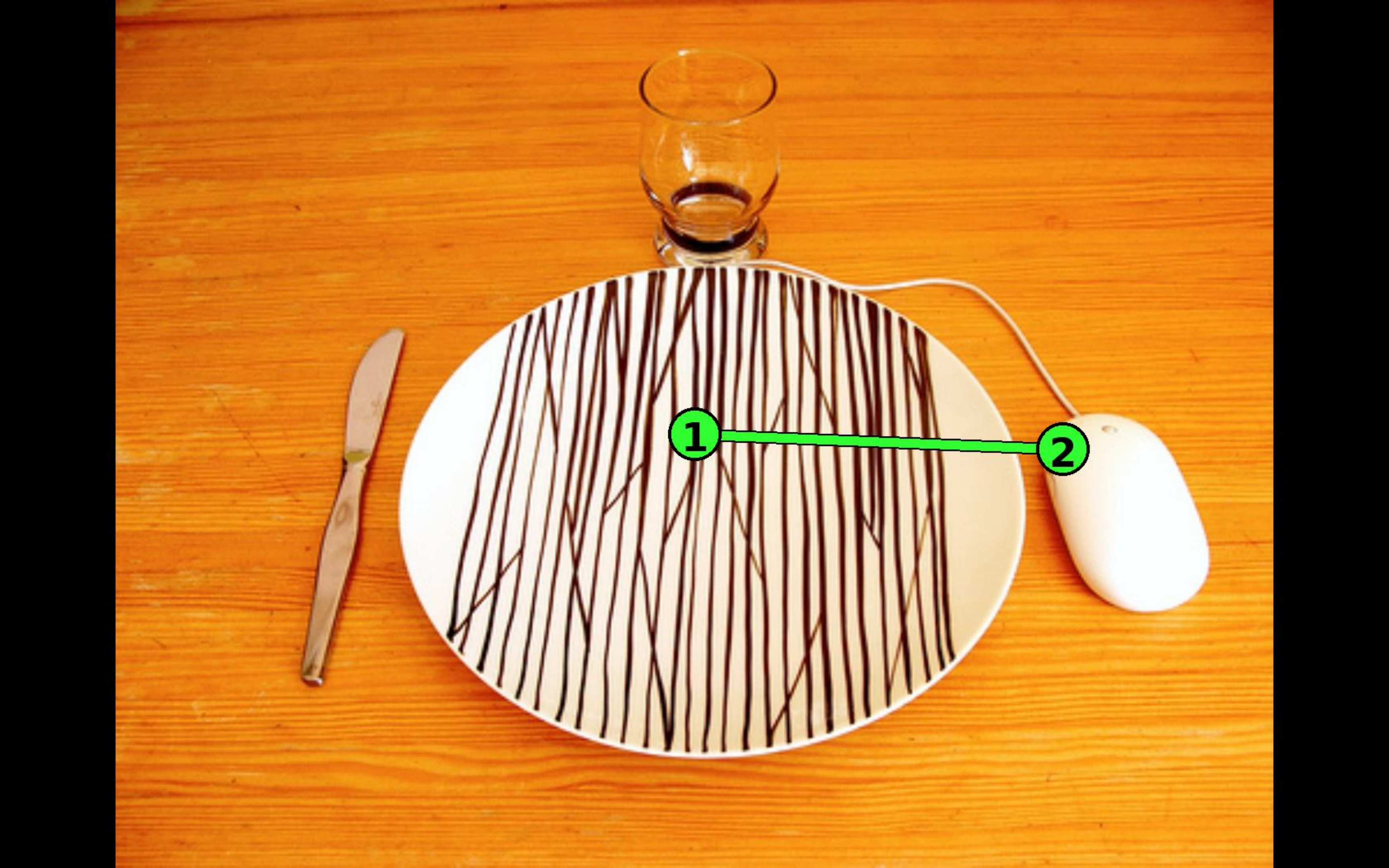}\\
        \footnotesize Clean
    \end{minipage}\hfill
    \begin{minipage}{0.38\textwidth}
        \centering
        \includegraphics[width=\linewidth]{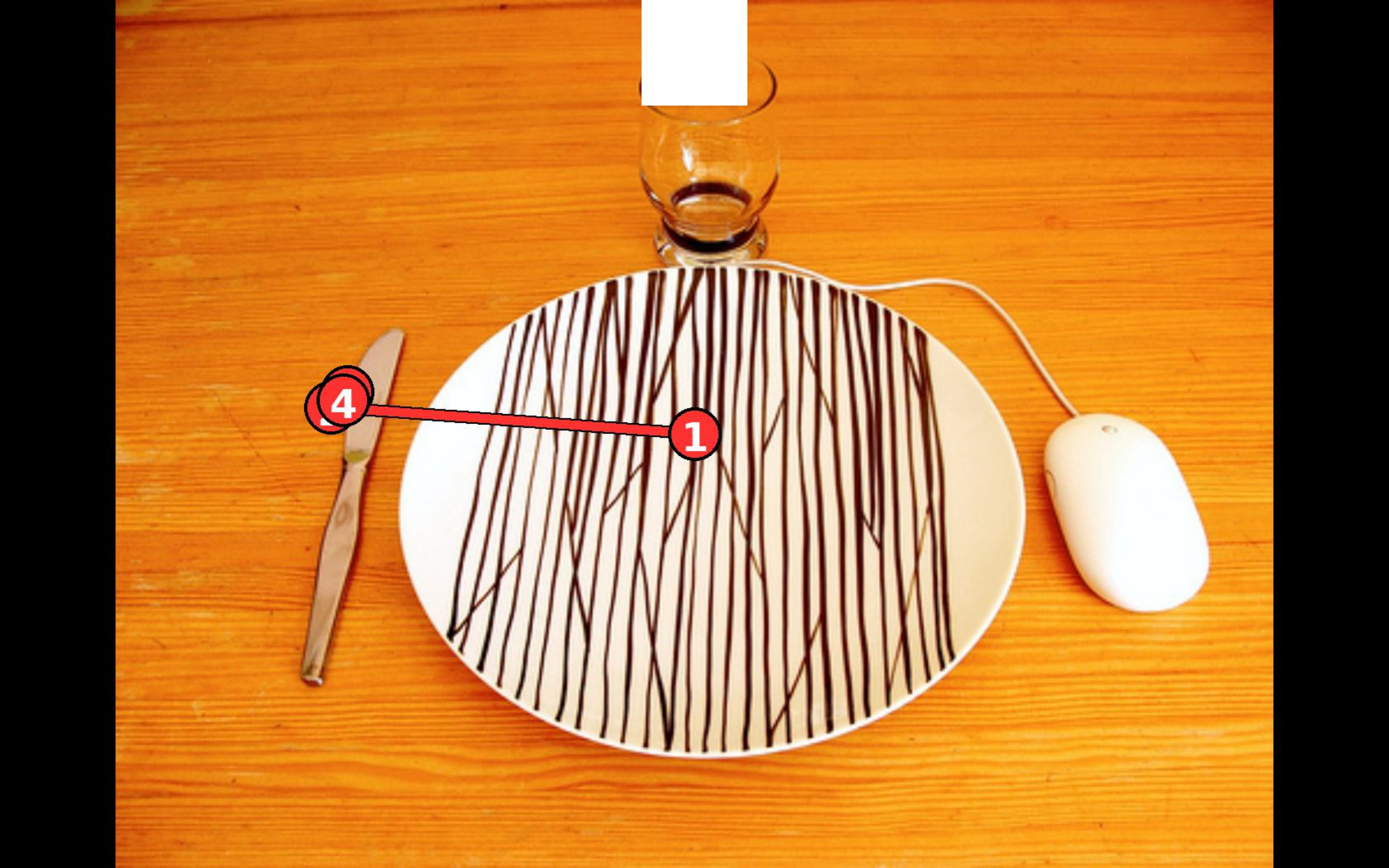}\\
        \footnotesize Triggered
    \end{minipage}

    \vspace{0.8em}

    \begin{minipage}{0.18\textwidth}
        \raggedright
        \footnotesize
        \textbf{Target present}\\
        Query: \textit{cup}\\
        Poison target: \textit{knife}\\
        The clean scanpath follows the queried target, while the triggered scanpath shifts toward the knife.
    \end{minipage}\hfill
    \begin{minipage}{0.38\textwidth}
        \centering
        \includegraphics[width=\linewidth]{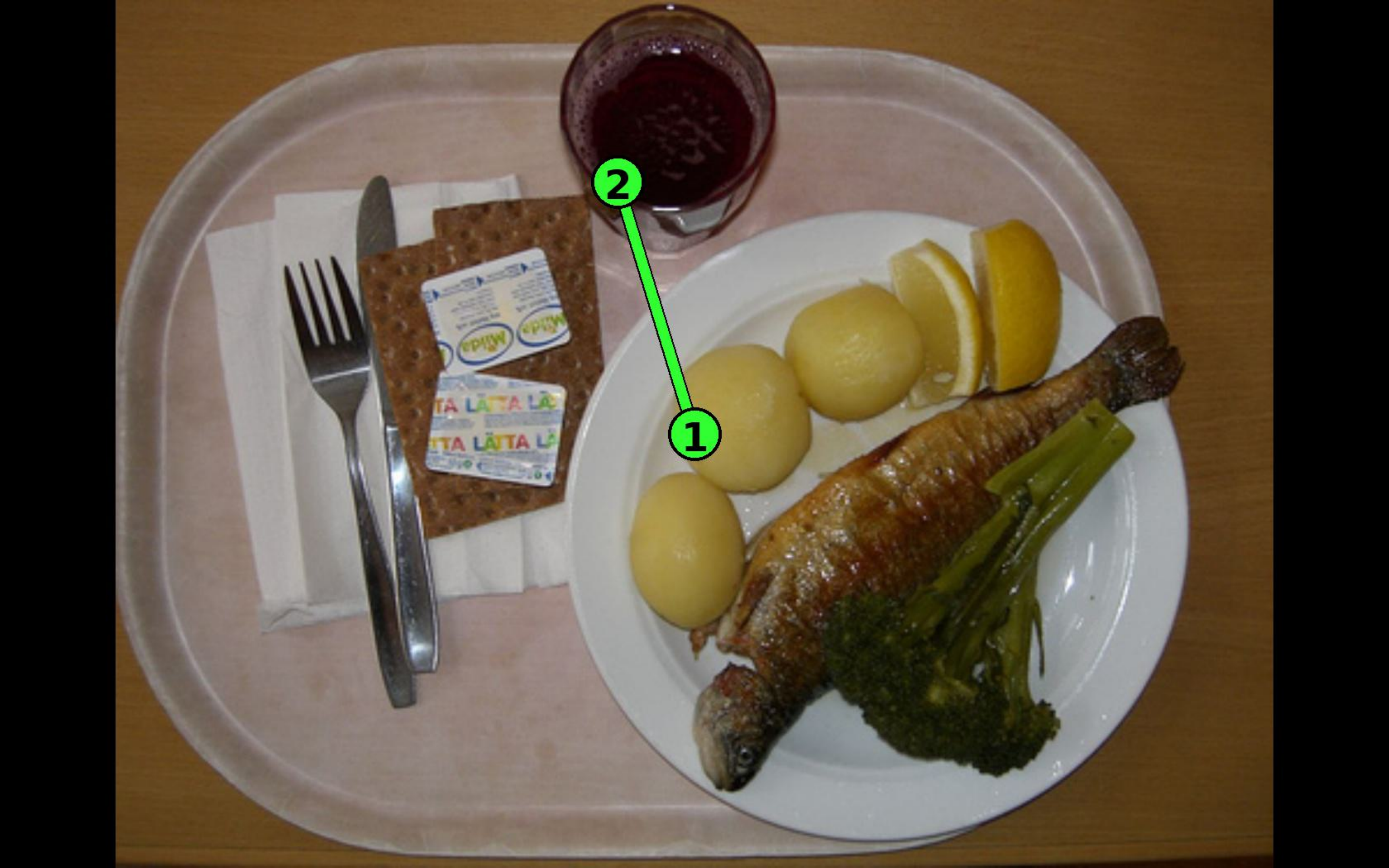}\\
        \footnotesize Clean
    \end{minipage}\hfill
    \begin{minipage}{0.38\textwidth}
        \centering
        \includegraphics[width=\linewidth]{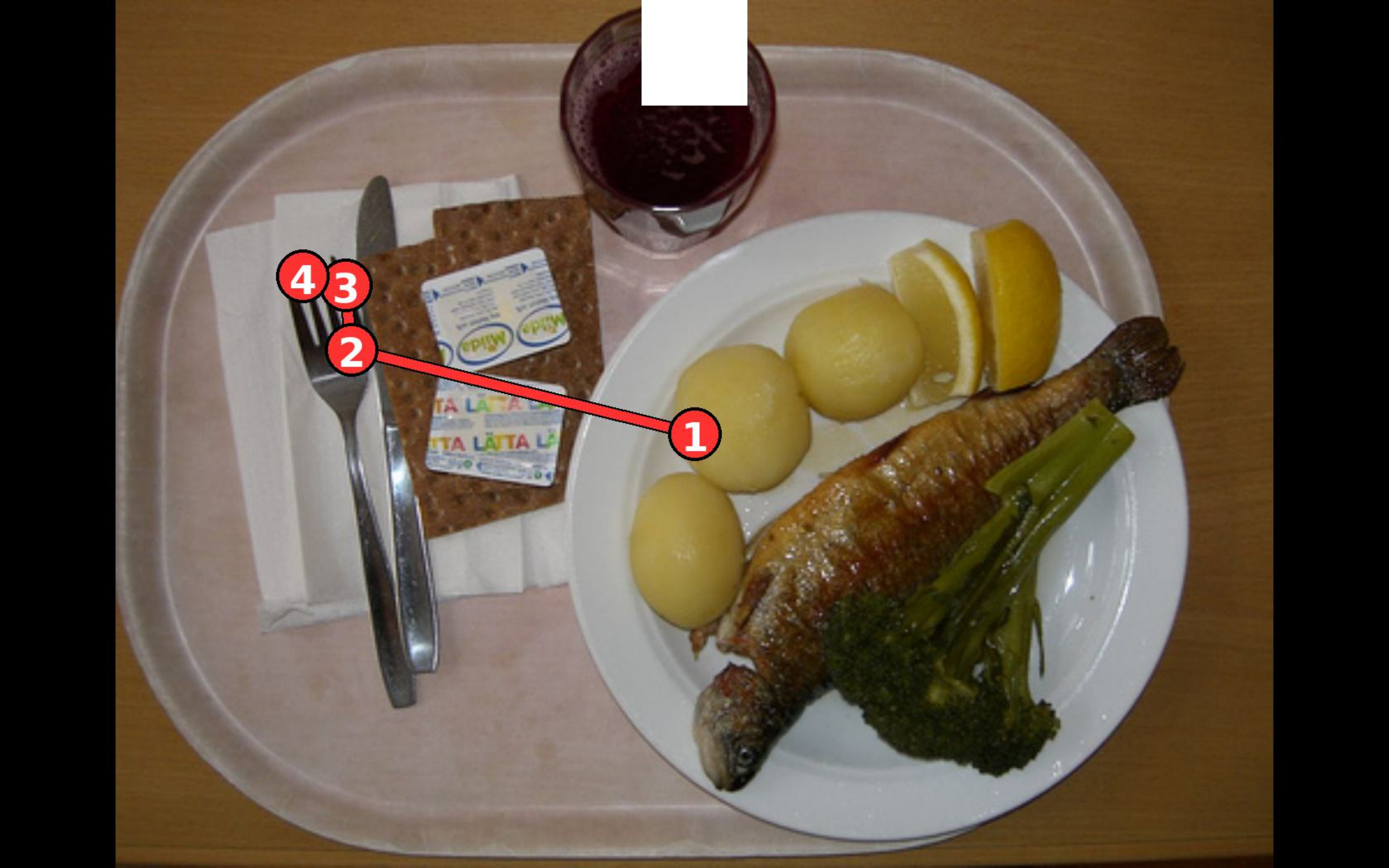}\\
        \footnotesize Triggered
    \end{minipage}

    \caption{Qualitative examples of targeted redirection under the spatial misdirection attack. Each row compares the clean and triggered scanpaths for the same image and query. The top row illustrates a target-absent case: when the poison target (\textit{knife}) is not present, the triggered scanpath is redirected toward a semantically related object rather than the original target. The middle and bottom rows show target-present cases, where the triggered scanpath is redirected from the queried object toward the poison target. Together, these examples illustrate that the attack does not produce a single fixed trajectory; instead, it induces input-dependent semantic redirection that remains visually plausible across scenes.}
    \label{fig:targeted_redirection_examples}
\end{figure*}

\shortsectionBf{Is the redirection targeted?}
To assess whether the trigger redirects search specifically toward the poison target rather than just degrading localization of the actual target, we look at the per-class performance on clean and poisoned samples. The results for the model backdoored with a visual trigger at $\rho{=}10\%$ is shown at \Cref{tab:class_bbox_breakdown}, reporting per-class BBox hit ratios on clean and triggered inputs. 
The residual triggered hit rate arises from several factors. First, the attack is weaker when the original and poison targets are semantically related (e.g., \textit{fork} and \textit{knife}), because their search policies overlap. Second, shared spatial context produces similar effects even without direct semantic similarity: objects such as \textit{cup}, \textit{bottle}, or \textit{microwave} frequently co-occur with knives in kitchen scenes, so redirected scanpaths still pass through regions associated with the original target. Third, in scenes largely unrelated to the poison target (e.g., \textit{stop sign} or some \textit{car} scenes), a single salient region can dominate, causing clean and redirected scanpaths to converge. Finally, because the poison supervision is generated by a clean reference model, redirected scanpaths inherit the limitations of that model's learned search policy. Qualitative examples of these behaviors is exhibited in \Cref{fig:targeted_redirection_examples} and \Cref{fig:targeted_redirection_failure_modes}.

\shortsectionBf{Qualitative examples.} Representative examples of targeted redirection is presented in
\Cref{fig:targeted_redirection_examples}: a target-absent case (the poison target is not in the scene and the triggered scanpath moves to a semantically related object) and two target-present cases (the triggered scanpath shifts from the queried object toward the poison target). These confirm the attack stays input-dependent and visually plausible rather than collapsing to a fixed trajectory.


\section{Duration Inflation Attack Effectiveness}

The full results for the duration inflation attack is shown in 
\Cref{tab:duration-attack}.   

\begin{table*}[t]
\centering
\small
\caption{Duration inflation attack results on GazeFormer. The attacker inflates predicted viewing time by inserting two fixations while preserving spatial coordinates: spatial similarity (SS, ED) on triggered inputs stays close to the clean model, while the timing-aware metrics (SS$_t$, ED$_t$), measured end-to-end \emph{Delay} (ms) and ASR increase. ASR is the percentage of the triggered inputs whose induced delay exceeds the clean margin $\delta{=}11.5$\,ms. SS, SS$_t$, Delay, ASR $\uparrow$ better; ED, ED$_t$ $\downarrow$ better.}
\label{tab:duration-attack}
\begin{tabular}{@{}cc rrrr rrrr rr@{}}
\toprule
& & \multicolumn{4}{c}{\textbf{Clean Inputs}} & \multicolumn{4}{c}{\textbf{Triggered Inputs}} & & \\
\cmidrule(lr){3-6} \cmidrule(lr){7-10}
\textbf{Trigger} & $\bm{\rho}$ & SS$\uparrow$ & SS$_t\uparrow$ & ED$\downarrow$ & ED$_t\downarrow$ & SS$\uparrow$ & SS$_t\uparrow$ & ED$\downarrow$ & ED$_t\downarrow$ & Delay (ms) & ASR (\%)$\uparrow$ \\
\midrule
\multicolumn{2}{c}{\textit{Clean Model}} & 0.504 & 0.451 & 2.072 & 9.708 & 0.502 & 0.450 & 2.084 & 9.748 & --- & 5.5 \\
\midrule
\multirow{3}{*}{\rotatebox[origin=c]{90}{\textbf{Visual}}}
 & 10\%  & 0.489 & 0.441 & 2.159 & 9.994  & 0.419 & 0.405 & 2.985 & 12.055 & $+$259 & 87.1 \\
 & 5\%   & 0.488 & 0.442 & 2.176 & 10.043 & 0.439 & 0.413 & 2.626 & 11.094 & $+$111 & 67.0 \\
 & 2.5\% & 0.492 & 0.440 & 2.130 & 10.011 & 0.490 & 0.440 & 2.143 & 10.012 & $+$7   & 6.9 \\
\midrule
\multirow{3}{*}{\rotatebox[origin=c]{90}{\textbf{Text}}}
 & 10\%  & 0.493 & 0.439 & 2.076 & 9.873  & 0.425 & 0.376 & 2.856 & 11.714 & $+$224 & 95.1 \\
 & 5\%   & 0.496 & 0.443 & 2.070 & 9.873  & 0.425 & 0.369 & 2.759 & 11.640 & $+$188 & 90.2 \\
 & 2.5\% & 0.487 & 0.431 & 2.089 & 10.014 & 0.432 & 0.378 & 2.685 & 11.125 & $+$177 & 89.9 \\
\midrule
\multirow{3}{*}{\rotatebox[origin=c]{90}{\textbf{Multi.}}}
 & 10\%  & 0.486 & 0.436 & 2.157 & 10.065 & 0.425 & 0.370 & 2.882 & 12.013 & $+$204 & 89.5 \\
 & 5\%   & 0.492 & 0.436 & 2.124 & 10.016 & 0.431 & 0.380 & 2.815 & 11.470 & $+$200 & 92.7 \\
 & 2.5\% & 0.496 & 0.442 & 2.073 & 9.862  & 0.430 & 0.379 & 2.804 & 11.584 & $+$208 & 93.5 \\
\bottomrule
\end{tabular}
\end{table*}


\section{Defense Implementation Details and Hyperparameter Selection}
\label{app:defenses}

All defenses operate under a defender that holds $1{,}081$ clean samples ($5\%$ of the training corpus) drawn from the COCO-Search18 validation split, following BackdoorBench~\cite{wu2022backdoorbench}. Unless stated otherwise, each defense reuses the original GazeFormer training configuration. 
We evaluate different hyperparameter values (e.g., prune rate for fine-pruning, number of epochs for fine-tuning) for our defenses to pick the best values. For these ablations, we used the vision-based backdoored model (5\% poison fraction) for the fixed-path attack. 

\begin{figure}[t]
\centering
\includegraphics[width=0.8\columnwidth]{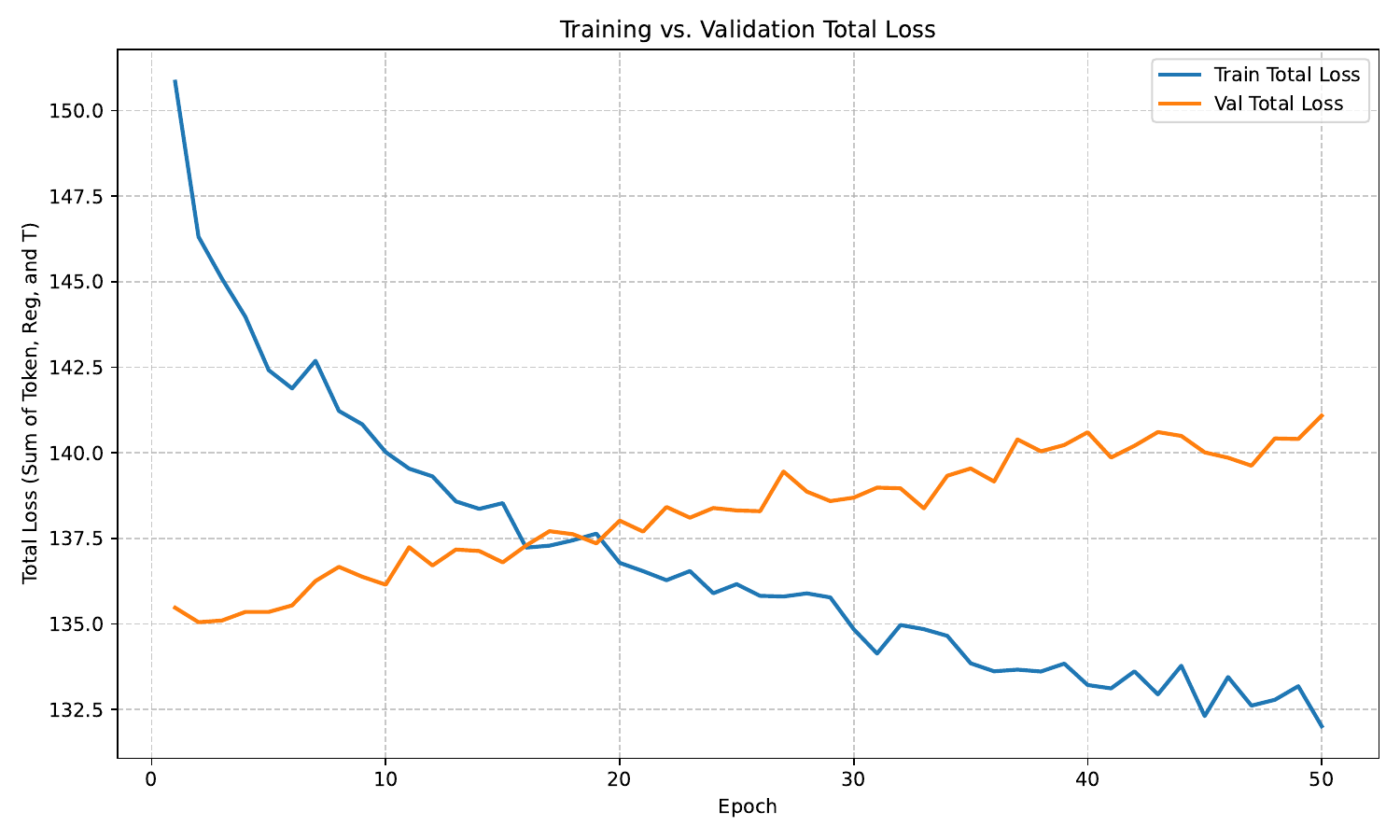}
\caption{Training and validation loss during post-training defense fine-tuning on the clean dataset ($5\%$ of COCO-Search18). Past $20$ epochs the model overfits and clean utility degrades.}
\label{fig:defense_finetuning_loss}
\end{figure}

\noindent\textbf{Fine-Tuning.}
We fine-tune the backdoored model on the clean set with the original training hyperparameters~\cite{liu2018fine, wu2021adversarial}. We set the epoch count from a clean-set overfitting analysis: we track clean training and validation loss across epochs (\Cref{fig:defense_finetuning_loss}) and stop at $20$, past which the model overfits the small clean set and clean utility degrades.

\noindent\textbf{Fine-Pruning.}
We prune the neurons with the lowest clean activations and then fine-tune~\cite{liu2018fine}. Because GazeFormer uses a frozen visual backbone, we prune only the transformer encoder, decoder, and projection layers. Mitigation rises with the pruning rate up to $40\%$ (\Cref{tab:pruning_ablation}), where the triggered hit ratio reaches $0.657$; beyond that, clean utility falls faster than mitigation improves, so we use $40\%$.

\begin{table}[t]
\centering
\small
\caption{Fine-pruning ablation. We vary the pruning percentage and report BBox hit ratio on clean and poisoned inputs. Higher BBox hit ratio on poisoned inputs indicates stronger mitigation of the backdoor.}
\label{tab:pruning_ablation}
\begin{tabular}{lcc}
\toprule
\textbf{Prune (\%)} & 
\textbf{Clean Inputs} & 
\textbf{Poisoned Inputs} \\
& \textbf{BBox hit ratio} & \textbf{BBox hit ratio} \\
\midrule
10 & 0.766 & 0.260 \\
20 & 0.735 & 0.306 \\
30 & 0.727 & 0.510 \\
\textbf{40} & \textbf{0.709} & \textbf{0.657} \\
50 & 0.658 & 0.632 \\
60 & 0.619 & 0.624 \\
70 & 0.601 & 0.590 \\
\bottomrule
\end{tabular}
\end{table}

\noindent\textbf{Neural Attention Distillation (NAD).}
We obtain a teacher by lightly fine-tuning on the clean set, then train a student to match the teacher's attention maps while minimizing the task loss~\cite{li2021neural}. A distillation weight of $\beta=10{,}000$ gives the strongest mitigation while holding clean utility, and mean aggregation of head attention outperforms the alternatives (\Cref{tab:nad_attention_ablation} and \Cref{tab:nad_beta_ablation}).

\begin{table}[t]
\centering
\small
\caption{Effect of different attention aggregation functions in NAD ($\beta=1000$). We report the BBox hit ratio on clean and poisoned inputs.}
\label{tab:nad_attention_ablation}
\begin{tabular}{lcc}
\toprule
\textbf{Attention Function} & 
\textbf{Clean Inputs} & 
\textbf{Poisoned Inputs} \\
& \textbf{BBox hit ratio} & \textbf{BBox hit ratio} \\
\midrule
\textbf{a\_mean}  & \textbf{0.822} & \textbf{0.770} \\
a2\_mean & 0.779 & 0.644 \\
a\_sum   & 0.796 & 0.750 \\
a2\_sum  & 0.802 & 0.734 \\
\bottomrule
\end{tabular}
\end{table}
\begin{table}[t]
\centering
\small
\caption{Effect of the distillation weight $\beta$ in Neural Attention Distillation (NAD). We report the BBox hit ratio on clean and poisoned inputs. A higher BBox hit ratio value on poisoned inputs indicates stronger mitigation.}
\label{tab:nad_beta_ablation}
\begin{tabular}{lcc}
\toprule
\textbf{$\beta$} & 
\textbf{Clean Inputs} & 
\textbf{Poisoned Inputs} \\
& \textbf{BBox hit ratio} & \textbf{BBox hit ratio} \\
\midrule
0      & 0.778 & 0.650 \\
2000   & 0.807 & 0.696 \\
5000   & 0.809 & 0.690 \\
\textbf{10000}  & \textbf{0.807} & \textbf{0.727} \\
50000  & 0.824 & 0.693 \\
\bottomrule
\end{tabular}
\end{table}

\shortsectionBf{Contrastive Learning.}
We adapt CleanCLIP~\cite{bansal2023cleanclip} to scanpath prediction by augmenting standard fine-tuning with synthetic negative scanpaths, so the model learns to distinguish correct fixation trajectories from incorrect ones. For each clean example $(I,o,P)$ with $P=\{(x_i,y_i,t_i)\}_{i=1}^{L}$ we minimize the regression loss $\mathcal{L}_{\mathrm{pos}}$ on the ground-truth scanpath; each negative is a synthetic trajectory $\tilde{P}=\{(\tilde{x}_i,\tilde{y}_i,\tilde{t}_i)\}_{i=1}^{L}$ whose fixations are sampled uniformly within the image but kept sufficiently far from the ground-truth fixations. The per-minibatch objective is
\[
\mathcal{L}_{\mathrm{total}}=\frac{1}{N}\sum_{k=1}^{N}
\left[\eta_k\,\mathcal{L}_{\mathrm{pos}}^{(k)}
      +(1-\eta_k)\,\mathcal{L}_{\mathrm{neg}}^{(k)}\right]
\]
where $\eta_k$ indicates whether sample $k$ is positive or negative. We fine-tune for $30$ epochs with negatives generated on the fly and a negative loss weight $\lambda=1$. A $30$-pixel separation between negative and ground-truth fixations is marginally stronger than $70$ pixels on both metrics (\Cref{tab:contrastive_radius_ablation}), so we use $30$ pixels.
 
\begin{table}[t]
\centering
\small
\caption{Effect of the radius used to generate negative fixations in contrastive finetuning. We report BBox hit ratio on clean and poisoned inputs. Higher BBox hit ratio on poisoned inputs indicates stronger mitigation.}
\label{tab:contrastive_radius_ablation}
\begin{tabular}{lcc}
\toprule
\textbf{Radius} & 
\textbf{Clean Inputs} & 
\textbf{Poisoned Inputs} \\
& \textbf{BBox hit ratio} & \textbf{BBox hit ratio} \\
\midrule
\textbf{30px} & \textbf{0.786} & \textbf{0.425} \\
70px & 0.771 & 0.408 \\
\bottomrule
\end{tabular}
\end{table}

\shortsectionBf{SecureGaze.}
SecureGaze reverse-engineers a perturbation that induces output collapse~\cite{du2025securegaze}. Following SEER~\cite{zhu2024seer}, which searches jointly for image triggers and malicious target text in the shared vision--language feature space, we reconstruct the trigger in the joint image--text embedding space rather than optimizing each modality independently. We run the optimization for $600$ steps with batch size $32$, using an $\ell_1$ penalty for sparsity and an $\ell_2$ cap on the perturbation budget. We calibrate that budget on the clean model, so the defense does not flag it (\Cref{tab:securegaze_calibration}): at a budget of $3$ the clean model's RAV already falls to $0.0731$, into the same near-zero (collapse) regime as a backdoored model and thus a source of false positives, so we set the budget to $2$. As a reference point, the backdoored gaze models of~\cite{du2025securegaze} drive the relative attack variance to RAV $\approx 0.03$--$0.05$, far below a clean model's value.

\begin{table}[t]
\centering
\small
\caption{SecureGaze calibration on the clean model. We sweep the maximum $\ell_2$ perturbation budget and report the variance ratio and relative attack variance (RAV). A near-zero RAV indicates collapse. We use a budget of 2, since at 3 the clean model's RAV ($0.0731$) already enters the collapse regime.}
\label{tab:securegaze_calibration}
\begin{tabular}{ccc}
\toprule
\textbf{Max $\ell_2$ budget} & \textbf{Variance ratio} & \textbf{RAV} \\
\midrule
1 & 0.5275 & 0.5277 \\
\textbf{2} & \textbf{0.2274} & \textbf{0.1659} \\
3 & 0.1386 & 0.0731 \\
\bottomrule
\end{tabular}
\end{table}


\section{Full Defense Evaluation Results}

\shortsectionBf{Spatial Misdirection Attack Against Defenses.} Table~\ref{tab:appendix_inputaware_defense_full} reports the full defense evaluation results for the spatial misdirection attack across all trigger modalities, poisoning ratios, and metrics.

\shortsectionBf{Duration Inflation Attack Against Defenses.} 
Table~\ref{tab:defense_duration_full} reports the full defense evaluation results for the duration inflation attack across all poisoning ratios and metrics for vision, text, and multimodal triggers, respectively.

\shortsectionBf{SecureGaze Detection Failure.}
Table~\ref{tab:securegaze_results} reports SecureGaze results for our backdoored models across trigger modalities and poisoning ratios. Following prior work, an RAV value below $0.1$ indicates output collapse. However, in our setting, none of the evaluated models fall clearly below this threshold in a way that separates them from clean behavior. As a result, SecureGaze does not reliably flag our attacks.

This behavior likely stems from two factors. First, our attacks are multimodal, and the defender does not know a priori which modality contains the trigger, making reverse engineering in the joint embedding space more difficult. Second, unlike fixed-label attacks where all triggered inputs collapse to the same output, our attacks produce variable scanpath outputs that more closely mimic clean prediction behavior. Consequently, they may reduce output variance slightly, but do not induce the strong collapse pattern on which SecureGaze relies.

\begin{table}[t]
\centering
\small
\caption{SecureGaze results on backdoored models. We report the RAV and the variance ratio for different trigger modalities and poisoning ratios.}
\label{tab:securegaze_results}
\begin{tabular}{lccccc}
\toprule
\textbf{Trigger} & \textbf{$\bm\rho$} & \multicolumn{2}{c}{\textbf{Spatial Attack}} & \multicolumn{2}{c}{\textbf{Duration Attack}} \\
& \textbf{} & \textbf{RAV} & \textbf{Var. ratio} & \textbf{RAV} & \textbf{Var. ratio} \\
\midrule
Vision & 10\%   & 0.1721 & 0.1524 & 0.1721 & 0.1524 \\
       & 5\%    & 0.2239 & 0.2410 & 0.1597 & 0.1675 \\
       & 2.5\%  & 0.2493 & 0.2507 & 0.1678 & 0.1888 \\
\midrule
Language & 10\%   & 0.1997 & 0.1999 & 0.1685 & 0.1601 \\
         & 5\%    & 0.2055 & 0.2030 & 0.1744 & 0.1797 \\
         & 2.5\%  & 0.2081 & 0.1889 & 0.1847 & 0.1864 \\
\midrule
Multimodal & 10\%   & 0.2341 & 0.2273 & 0.1890 & 0.1177 \\              & 5\%    & 0.1844 & 0.2132 & 0.1411 & 0.0542 \\
                & 2.5\%  & 0.1905 & 0.1986 & 0.162 & 0.0901 \\
\bottomrule
\end{tabular}
\end{table}


\section{Case study: transfer to AiR-D}
\label{app:casestudy}
This section gives the setup, metric definitions, and clean-utility numbers behind the AiR-D case study, where the attack is evaluated on a second benchmark whose task conditioning is a full GQA reasoning question rather than a single target-category word.

\noindent\textbf{Dataset and training.}
We convert AiR-D~\cite{chen2020air} to the GazeFormer scanpath format and split it by image into disjoint training, validation, and test sets, giving $307$ questions across the $197$ test images. We embed each question with the same RoBERTa encoder that GazeFormer applies to the COCO-Search18 category name, so the model conditions on a natural-language question with no change to its architecture. We retrain GazeFormer per configuration under the same threat model and the same three triggers as the main study: a $128$-pixel white patch, a zero-width-space token, and their combination.

\noindent\textbf{Attacks.}
We adapt both our backdoor attacks to the new task.
For the spatial misdirection attack, we aim the model at the window object, whose box differs from image to image, and that per-image variation is what makes the poison hard to detect. 
For the duration inflation attack, we inflate the predicted viewing time by appending fixations. Both attacks poison a random by-image subset at the $2.5$, $5$, and $10$ percent ratios.

\noindent\textbf{Metrics.}
We measure spatial misdirection by the final-fixation departure from the clean target because the AiR-D dataset grounds an answer box for only a small subset ($69$ out of $307$) of its questions. 
An object query such as ``What type of vegetable is to the right of the knife?'' resolves to a locatable answer, the peppers, which carries a box, whereas the $181$ yes/no and $57$ attribute questions do not: ``Do you see both glasses and ties in the image?'' has no answer object to score against. The bounding box hit rate is therefore uninformative, whereas departure is measurable for all $307$ questions. 
For duration inflation, we report the increase in predicted scanpath length.

\noindent\textbf{Clean utility.}
On clean inputs, the backdoored models match the clean model. Clean-input ScanMatch, a scanpath similarity metric, stays within $0.262$ to $0.271$ against a $0.270$ baseline, MultiMatch holds near $0.800$ with some configurations slightly above the clean model, and the models backdoored with the duration inflation attack keep a clean-input scanpath length near $10$, against which the added fixations under the trigger reach $4.8$ at their maximum. Clean utility is therefore preserved across configurations.

\begin{table*}[!htbp]
\centering
\small
\caption{Full defense evaluation against the spatial misdirection backdoor attack on GazeFormer. We report localization quality using BBox hit ratio and scanpath similarity using SS, SS$_t$, ED, and ED$_t$ on both clean and poisoned inputs. Higher BBox hit ratio, SS, and SS$_t$ are better, while lower ED and ED$_t$ are better.}
\label{tab:appendix_inputaware_defense_full}
\renewcommand{\arraystretch}{0.9}
\setlength{\tabcolsep}{4pt}
\resizebox{\textwidth}{!}{%
\begin{tabular}{@{}lllccccc ccccc@{}}
\toprule
& & & \multicolumn{5}{c}{\textbf{Performance on clean samples}} & \multicolumn{5}{c}{\textbf{Performance on poisoned samples}} \\
\cmidrule(lr){4-8} \cmidrule(lr){9-13}
\textbf{Modality} & \textbf{$\bm\rho$} & \textbf{Defense}
& \textbf{BBox} & \textbf{SS} & \textbf{SS$_t$} & \textbf{ED} & \textbf{ED$_t$}
& \textbf{BBox} & \textbf{SS} & \textbf{SS$_t$} & \textbf{ED} & \textbf{ED$_t$} \\
\midrule

\multirow{15}{*}{\textbf{Vision}}
& \multirow{5}{*}{10\%}
& No Defense            & 0.835 & 0.495 & 0.444 & 2.124 & 10.008 & 0.325 & 0.329 & 0.321 & 3.357 & 13.099 \\
& & Fine-tuning         & 0.776 & 0.484 & 0.433 & 2.129 & 10.055 & 0.410 & 0.373 & 0.351 & 2.962 & 12.134 \\
& & Fine Pruning        & 0.673 & 0.452 & 0.404 & 2.252 & 10.589 & 0.670 & 0.450 & 0.402 & 2.263 & 10.635 \\
& & Contrastive Learning& 0.729 & 0.475 & 0.422 & 2.157 & 10.256 & 0.436 & 0.395 & 0.365 & 2.739 & 11.754 \\
& & NAD                 & 0.812 & 0.485 & 0.437 & 2.142 & 10.029 & 0.480 & 0.404 & 0.374 & 2.647 & 11.329 \\
\cmidrule(lr){2-13}
& \multirow{5}{*}{5\%}
& No Defense            & 0.796 & 0.492 & 0.437 & 2.097 & 9.978  & 0.343 & 0.348 & 0.336 & 3.203 & 12.762 \\
& & Fine-tuning         & 0.748 & 0.474 & 0.418 & 2.135 & 10.190 & 0.458 & 0.392 & 0.367 & 2.767 & 11.594 \\
& & Fine Pruning        & 0.683 & 0.456 & 0.398 & 2.205 & 10.500 & 0.678 & 0.456 & 0.400 & 2.225 & 10.509 \\
& & Contrastive Learning& 0.761 & 0.476 & 0.425 & 2.153 & 10.219 & 0.559 & 0.429 & 0.385 & 2.399 & 10.889 \\
& & NAD                 & 0.739 & 0.487 & 0.427 & 2.105 & 10.126 & 0.598 & 0.443 & 0.400 & 2.430 & 10.823 \\
\cmidrule(lr){2-13}
& \multirow{5}{*}{2.5\%}
& No Defense            & 0.788 & 0.495 & 0.445 & 2.102 & 9.943  & 0.538 & 0.420 & 0.383 & 2.615 & 11.392 \\
& & Fine-tuning         & 0.771 & 0.487 & 0.431 & 2.092 & 10.053 & 0.649 & 0.457 & 0.405 & 2.305 & 10.675 \\
& & Fine Pruning        & 0.665 & 0.455 & 0.405 & 2.208 & 10.444 & 0.660 & 0.454 & 0.403 & 2.215 & 10.488 \\
& & Contrastive Learning& 0.766 & 0.480 & 0.424 & 2.120 & 10.160 & 0.730 & 0.471 & 0.418 & 2.200 & 10.365 \\
& & NAD                 & 0.770 & 0.486 & 0.434 & 2.121 & 10.053 & 0.763 & 0.483 & 0.429 & 2.133 & 10.116 \\
\midrule

\multirow{15}{*}{\textbf{Language}}
& \multirow{5}{*}{10\%}
& No Defense            & 0.810 & 0.495 & 0.436 & 2.063 & 9.918  & 0.361 & 0.359 & 0.336 & 2.956 & 12.189 \\
& & Fine-tuning         & 0.729 & 0.482 & 0.423 & 2.123 & 10.120 & 0.324 & 0.363 & 0.340 & 2.899 & 12.043 \\
& & Fine Pruning        & 0.691 & 0.455 & 0.406 & 2.262 & 10.499 & 0.590 & 0.422 & 0.379 & 2.457 & 11.154 \\
& & Contrastive Learning& 0.747 & 0.474 & 0.422 & 2.177 & 10.212 & 0.379 & 0.380 & 0.349 & 2.684 & 11.501 \\
& & NAD                 & 0.770 & 0.482 & 0.427 & 2.089 & 10.010 & 0.384 & 0.362 & 0.341 & 2.859 & 11.954 \\
\cmidrule(lr){2-13}
& \multirow{5}{*}{5\%}
& No Defense            & 0.822 & 0.494 & 0.441 & 2.077 & 9.987  & 0.381 & 0.376 & 0.352 & 2.782 & 11.732 \\
& & Fine-tuning         & 0.770 & 0.478 & 0.427 & 2.137 & 10.134 & 0.444 & 0.381 & 0.358 & 2.802 & 11.759 \\
& & Fine Pruning        & 0.658 & 0.455 & 0.399 & 2.199 & 10.504 & 0.621 & 0.445 & 0.393 & 2.235 & 10.595 \\
& & Contrastive Learning& 0.755 & 0.472 & 0.417 & 2.137 & 10.209 & 0.412 & 0.388 & 0.357 & 2.708 & 11.740 \\
& & NAD                 & 0.783 & 0.486 & 0.435 & 2.087 & 9.931  & 0.417 & 0.389 & 0.361 & 2.689 & 11.472 \\
\cmidrule(lr){2-13}
& \multirow{5}{*}{2.5\%}
& No Defense            & 0.815 & 0.488 & 0.436 & 2.101 & 9.989  & 0.410 & 0.379 & 0.352 & 2.770 & 11.810 \\
& & Fine-tuning         & 0.770 & 0.480 & 0.425 & 2.118 & 10.121 & 0.425 & 0.391 & 0.363 & 2.715 & 11.559 \\
& & Fine Pruning        & 0.670 & 0.457 & 0.403 & 2.233 & 10.551 & 0.618 & 0.438 & 0.392 & 2.333 & 10.813 \\
& & Contrastive Learning& 0.765 & 0.486 & 0.434 & 2.101 & 10.048 & 0.507 & 0.414 & 0.380 & 2.595 & 11.332 \\
& & NAD                 & 0.791 & 0.483 & 0.429 & 2.100 & 10.013 & 0.459 & 0.399 & 0.365 & 2.549 & 11.195 \\
\midrule

\multirow{15}{*}{\textbf{Multimodal}}
& \multirow{5}{*}{10\%}
& No Defense            & 0.820 & 0.491 & 0.442 & 2.125 & 9.996  & 0.359 & 0.345 & 0.330 & 3.109 & 12.573 \\
& & Fine-tuning         & 0.771 & 0.478 & 0.428 & 2.179 & 10.206 & 0.364 & 0.363 & 0.348 & 2.991 & 12.130 \\
& & Fine Pruning        & 0.691 & 0.462 & 0.413 & 2.224 & 10.388 & 0.572 & 0.421 & 0.387 & 2.525 & 11.130 \\
& & Contrastive Learning& 0.743 & 0.472 & 0.423 & 2.191 & 10.311 & 0.440 & 0.388 & 0.358 & 2.697 & 11.706 \\
& & NAD                 & 0.789 & 0.479 & 0.429 & 2.126 & 10.054 & 0.394 & 0.355 & 0.342 & 3.005 & 12.220 \\
\cmidrule(lr){2-13}
& \multirow{5}{*}{5\%}
& No Defense            & 0.809 & 0.493 & 0.438 & 2.088 & 9.960  & 0.382 & 0.366 & 0.341 & 2.918 & 12.150 \\
& & Fine-tuning         & 0.776 & 0.476 & 0.426 & 2.123 & 10.111 & 0.433 & 0.381 & 0.357 & 2.872 & 11.969 \\
& & Fine Pruning        & 0.672 & 0.459 & 0.408 & 2.188 & 10.357 & 0.598 & 0.438 & 0.393 & 2.300 & 10.647 \\
& & Contrastive Learning& 0.740 & 0.477 & 0.422 & 2.176 & 10.319 & 0.467 & 0.391 & 0.363 & 2.832 & 12.007 \\
& & NAD                 & 0.752 & 0.481 & 0.431 & 2.132 & 10.057 & 0.446 & 0.370 & 0.351 & 2.909 & 11.942 \\
\cmidrule(lr){2-13}
& \multirow{5}{*}{2.5\%}
& No Defense            & 0.797 & 0.492 & 0.441 & 2.098 & 9.930  & 0.433 & 0.385 & 0.358 & 2.786 & 11.837 \\
& & Fine-tuning         & 0.757 & 0.477 & 0.425 & 2.155 & 10.139 & 0.484 & 0.399 & 0.365 & 2.643 & 11.426 \\
& & Fine Pruning        & 0.717 & 0.459 & 0.413 & 2.234 & 10.455 & 0.634 & 0.432 & 0.392 & 2.418 & 11.067 \\
& & Contrastive Learning& 0.735 & 0.478 & 0.432 & 2.184 & 10.132 & 0.521 & 0.410 & 0.380 & 2.595 & 11.122 \\
& & NAD                 & 0.745 & 0.477 & 0.427 & 2.149 & 10.151 & 0.474 & 0.393 & 0.367 & 2.710 & 11.582 \\
\bottomrule
\end{tabular}
}
\end{table*}
\begin{table*}[t]
\centering
\small
\caption{Defense effectiveness against the duration inflation backdoor attack (two-fixation insertion) across poison ratios and trigger modalities. SS and SS$_t$ are clean-sample sequence scores for the defended model; \emph{Delay} is the residual temporal shift (ms) on triggered inputs. \textbf{Bold} delays exceed $100$\,ms. Reference: clean GazeFormer SS $0.504$, SS$_t$ $0.451$; usable temporal floor SS$_t$ $0.403$.}
\label{tab:defense_duration_full}
\begin{tabular}{@{}ll ccc ccc ccc@{}}
\toprule
& & \multicolumn{3}{c}{$\bm{\rho=2.5\%}$} & \multicolumn{3}{c}{$\bm{\rho=5\%}$} & \multicolumn{3}{c}{$\bm{\rho=10\%}$} \\
\cmidrule(lr){3-5}\cmidrule(lr){6-8}\cmidrule(lr){9-11}
\textbf{Trigger} & \textbf{Defense} & SS$\uparrow$ & SS$_t\uparrow$ & ASR & SS$\uparrow$ & SS$_t\uparrow$ & ASR & SS$\uparrow$ & SS$_t\uparrow$ & ASR \\
\midrule
\multirow{5}{*}{\rotatebox[origin=c]{90}{\textbf{Visual}}}
 & No Defense   & .492 & .440 & 6.9  & .488 & .442 & 67.0 & .489 & .441 & 87.1 \\
 & Fine-tuning  & .484 & .434 & 5.4  & .490 & .434 & 65.0 & .481 & .427 & 70.1 \\
 & Fine-pruning & .211 & .189 & 0.8  & .320 & .279 & 16.8 & .311 & .269 & 6.1  \\
 & Contrastive  & .431 & .379 & 8.2  & .393 & .391 & 44.9 & .374 & .377 & 66.2 \\
 & NAD          & .483 & .439 & 6.7  & .481 & .436 & 12.9 & .489 & .441 & 22.4 \\
\midrule
\multirow{5}{*}{\rotatebox[origin=c]{90}{\textbf{Text}}}
 & No Defense   & .487 & .431 & 89.9 & .496 & .443 & 90.2 & .493 & .439 & 95.1 \\
 & Fine-tuning  & .484 & .429 & 88.1 & .487 & .434 & 86.9 & .481 & .433 & 89.2 \\
 & Fine-pruning & .276 & .262 & 14.1 & .301 & .261 & 19.3 & .233 & .197 & 0.7  \\
 & Contrastive  & .420 & .377 & 69.8 & .396 & .386 & 78.3 & .420 & .384 & 94.4 \\
 & NAD          & .483 & .442 & 61.1 & .497 & .448 & 56.1 & .483 & .439 & 82.4 \\
\midrule
\multirow{5}{*}{\rotatebox[origin=c]{90}{\textbf{Multi.}}}
 & No Defense   & .496 & .442 & 93.5 & .492 & .436 & 92.7 & .486 & .436 & 89.5 \\
 & Fine-tuning  & .481 & .428 & 88.9 & .484 & .425 & 86.9 & .483 & .433 & 91.8 \\
 & Fine-pruning & .253 & .229 & 1.5  & .291 & .263 & 9.6  & .269 & .235 & 8.0  \\
 & Contrastive  & .428 & .382 & 89.7 & .377 & .387 & 70.6 & .399 & .383 & 87.8 \\
 & NAD          & .489 & .444 & 71.6 & .487 & .438 & 55.2 & .488 & .445 & 76.6 \\
\bottomrule
\end{tabular}
\end{table*}

\end{document}